\documentstyle[epsf,twocolumn,floats,prb,aps]{revtex}
\begin{document}
\draft

 \twocolumn[\hsize\textwidth\columnwidth\hsize\csname @twocolumnfalse\endcsname

\title{
Green's Function Monte Carlo for Lattice Fermions:\\
Application to the {\em t-J} Model}
\author{ C. Stephen Hellberg$^{\dagger}$ and Efstratios Manousakis$^{\ddagger}$ }
\address{ $^{\dagger}$Center for Computational Materials Science, Naval Research Laboratory, Washington, D.C., 20305\\
$^{\ddagger}$Department of Physics and MARTECH,
	Florida State University, Tallahassee, FL 32306-3016 }
\date{\today}
\maketitle
\begin{abstract}
\noindent
We develop a general numerical method to study the zero temperature
properties of strongly correlated electron models on large lattices.
The technique, which resembles Green's Function Monte Carlo,
projects the ground state component from a trial wave function with
no approximations.
We use this method
to determine the phase diagram of the two-dimensional
{\em t-J} model, using the Maxwell construction to investigate electronic
phase separation.
The shell effects of fermions on finite-sized periodic lattices are
minimized by keeping the number of electrons fixed at a closed-shell 
configuration and varying the size of the lattice.
Results obtained for various electron
numbers corresponding to different closed-shells indicate that the
finite-size effects in our calculation are small. 
For any value of interaction strength,
we find that there is always a value
of the electron density 
above which the
system can lower its energy by forming a two-component phase separated state.
Our results are compared with other
calculations on the {\em t-J} model. We find that the most accurate results
are consistent with phase separation at all interaction strengths.

\end{abstract}

 ]

\section{Introduction}

Correlated quantum many-body systems have
provided a host of new phenomena such
as new states of matter, new forms of ordering
transitions, etc. These phenomena are the result of the
appearance of fundamentally new relevant degrees of
freedom which emerge as a coherent superposition of
the underlying degrees of freedom of the many variable system.
Once one 
identifies the important degrees of freedom,
these degrees of freedom can be
treated by analytical or semi-analytical techniques 
which are variants of generalized perturbation expansions
around the defining framework. The problem which arises
is that such frameworks cannot be imagined before hand unless
there are hints from either experiment or numerical
studies of models correctly capturing the dynamics
of more basic degrees of freedom which, at
first sight, seem featureless. 

Several such models on a discrete lattice
exist and a variety of numerical techniques are
at our disposal to use.\cite{manousakis91}
Exact diagonalization
techniques suffer from the fact that
the dimensionality $N_H$ of the Hilbert space 
grows exponentially with 
system size $N_s$ (number of sites). Taking into account all the symmetries
of the problem can reduce the size of the
invariant subspaces to smaller size $N_R$ (which may be a few orders of
magnitude smaller than $N_H$). However, the largest possible size
increases only with the logarithm of the ratio of $N_H/N_R$.
In particular, most interesting quantities scale with the
linear dimension of the system which scales with $[ln(N_H/N_R)]^{1\over d}$ where
$d$ is the dimensionality of the problem. Renormalization group approaches, 
such as the density matrix renormalization group (DMRG) technique,\cite{white93} have been very 
successful in one dimension but there are significant limitations in 
higher dimensions.

Attractive alternatives seem to be
stochastic methods such as quantum Monte Carlo which can give
information about larger size systems.
Many interesting problems,
however, involve fermionic degrees of freedom.
If one attempts  a simulation of fermions at low temperature 
one encounters the so-called fermion sign problem. Namely,
one needs to define configurations which carry a phase
(a positive or negative sign) along with their statistical
weights, a reflection of the transformation property of the fermion
wave function under particle permutation. In the computation of
many quantities of interest, such as the energy, the ``positive''
and the ``negative'' configurations give nearly opposite contributions,
leading to 
wildly fluctuating weights.
The negative-sign problem causes the statistical fluctuations to diverge 
exponentially with increasing system size for fixed density.

The Green's function Monte Carlo (GFMC) method has been successfully 
applied to lattice spin systems, in particular
to the square lattice spin-1/2 Heisenberg quantum 
antiferromagnet.\cite{manousakis91,runge92,trivedi89,trivedi90,carlson89,gross89}
In this case, through the Marshall-sign transformation, the problem
can be mapped to a hard-core boson problem which presents no
sign problem and solved accurately on large size systems. 

An approximate method to deal with the sign problem in fermionic systems
is the fixed node (FN)
approximation.\cite{ceperley80,ceperley86}
This approach
projects a trial state onto the best variational state
with the same nodal structure, thus controlling the statistical fluctuations.
The FN approximation has been used for lattice fermion systems also.\cite{vanBemmel94,tenHaaf95a,tenHaaf95b}

The GFMC method for lattice fermions without the fixed node approximation has been applied 
to one-dimension systems.\cite{hellberg93}
In two dimensions it has been applied to the
\mbox{\em t-J} model in the limits
of small numbers of electrons\cite{hellberg95} 
or holes.\cite{boninsegni93,boninsegni92}

In this paper, 
we present an efficient implementation of GFMC for lattice fermions
at arbitrary densities of electrons or holes.
We demonstrate the
utility of the method in the case of the two dimensional \mbox{\em t-J} model.
The method projects a trial wave function onto the lowest energy
eigenstate that it overlaps.
If the trial state overlaps the ground state, the projection yields the
ground state.
The projection becomes statistically more accurate as the ground-state
component of the trial state increases relative to the excited-state
components. Results obtained for the \mbox{\em t-J} model with this method 
have been published by the authors.\cite{hellberg97}
In this paper we present the general method and in addition, our results
for the \mbox{\em t-J} model are presented in detail and compared with other
recent calculations.

The \mbox{\em t-J} model is thought to contain some important aspects 
of the environment in the copper-oxide superconductors. For instance,  
the calculated single-hole spectrum\cite{liu} is in agreement 
with the results 
of the photo-emission data.\cite{wells}
The model gives rise to 
a two-hole bound state\cite{boninsegni93} with the $d_{x^2-y^2}$ 
symmetry which is the believed symmetry of the superconducting 
state in these materials.
In addition, Emery, Kivelson, and Lin (EKL)\cite{emery90} suggested that the cuprates are near 
an electronic phase separation (PS) instability which is prevented by the 
long-range part of the Coulomb interaction.
In the phase-separated state, the holes cluster together with a certain
density of electrons,
leaving the rest of the system in an antiferromagnetic state with no holes.
Phase separation in the \mbox{\em t-J} model
has been studied by a number of techniques which seem to be giving conflicting
conclusions.\cite{hellberg97,hellberg98,dagotto94,luchini91,putikka92,poilblanc95,kohno,sorella98}
High temperature series expansions\cite{luchini91,putikka92}
and some studies on small systems\cite{poilblanc95} indicated that
phase separation does not occur at the physical region for the cuprates,
namely $J/t \sim 0.3-0.4$,
while other studies on small systems\cite{emery90,hellberg98}
found this region was unstable to phase separation.
Using the GFMC method presented in this paper,
Hellberg and Manousakis\cite{hellberg97}
found that the \mbox{\em t-J} model has a region of phase 
separation at {\em all} interaction strengths.  

In recent work,
Calandra, Becca, and Sorella\cite{sorella98} (CBS) emphasize 
that the phase separated region does not extend below
$J/t \lesssim 0.4$.
In addition, the DMRG method has been also applied to this problem
by White and Scalapino (WS)\cite{white98a,white98b} who find that the ground state of the
\mbox{\em t-J} model on the square lattice is characterized by stripes.

In the last section of this paper, we compare our
results to those of CBS and WS,
concluding that
the physical region of the model is very close to or inside of
a PS instability. Namely, the early conclusions
that {\it the physical region of the \mbox{t-J} model is far from
the critical $J_c$ for phase separation}, are largely invalid.

\section{Numerical Method}

Even though our formalism is general and can be applied to 
other lattice fermion models, we shall use the example of
the \mbox{\em t-J} Hamiltonian on a two-dimensional square lattice. 
This model in itself is a non-trivial extension of 
the square lattice Heisenberg antiferromagnet\cite{manousakis91}
where GFMC was first applied on a lattice model.

The {\em t-J} model  is written in the subspace with no 
doubly occupied sites as
\begin{equation}
   H
   = - t
      \sum_{\langle ij \rangle \sigma}
      ( c_{i\sigma}^{\dagger}c_{j\sigma}
		+ {\rm h.c.}
		)
 +
      J
\sum_{\langle ij \rangle}
      ( {\bf S}_{i} \! \cdot \! {\bf S}_{j} -
         \frac{n_i n_j}{4} ) .
\label{tj-ham}
\end{equation}
Here $\langle ij \rangle$ enumerates neighboring sites
on a square lattice,
$c_{i\sigma}^{\dagger}$ creates an electron of spin $\sigma$ on site
$i$, $n_i = \sum_\sigma c_{i\sigma}^{\dagger}c_{i\sigma}$,
and ${\bf S}_{i}$ is the spin-$\frac{1}{2}$ operator.

\subsection{Details of the Projection}

We take a trial wave function $\Psi$ and project
it onto the ground state by generating a series of increasingly accurate
approximants to the ground state labeled by integers
$ |m \rangle = (H - W)^m | \Psi\rangle$.
Here $H$ is the Hamiltonian and $W$ is an appropriately chosen numerical
constant.
\cite{liang90,liang90.2,chen93,chen93.2}

We may expand the trial state in terms of the exact eigenstates:
\begin{equation}
|\Psi\rangle = a_0 | \Phi_0 \rangle + a_1 | \Phi_1 \rangle + \cdots
\end{equation}
where $ | \Phi_0 \rangle$ is the ground state, $| \Phi_1 \rangle$ is the
first excited state, and the $a_i$'s are the expansion coefficients.
Rewriting the projected states in this way, we see
\begin{eqnarray}
   | m \rangle & = & ( H - W )^m | \Psi \rangle\\
   & = & a_0 (E_0  - W )^m| \Phi_0 \rangle
		+ a_1  (E_1  - W )^m | \Phi_1 \rangle + \cdots\\
   & \propto & 
   \left\{ | \Phi_0 \rangle + \frac{a_1}{a_0} \left( \frac{ E_1 - W }{ E_0 - W}
    \right) ^m  | \Phi_1 \rangle + \cdots \right\}
\label{expand}
\end{eqnarray}
where $E_i$ is the energy of the $i$'th eigenstate.
So $| m \rangle$ approaches the ground state
for large $m$ provided
\begin{equation}
| E_{i>0} - W | < |E_0 - W | \label{project:cond}
\end{equation}
for all excited state energies $E_{i>0}$.
The projection can be formulated in this simple way because eigenvalues
of lattice Hamiltonians are bounded from below {\em and} above.
Continuum problems require a different form for the projection operator.
\cite{li91,caffarel92}
In what follows, we assume the offset constant $W$ is incorporated in the
Hamiltonian.

From (\ref{expand}) we see rate of convergence with $m$
is governed by the overlap of the trial state with the ground state
and the energy of the lowest excited state overlapping the trial state.
In Section \ref{trial_section}, we describe the steps taken to insure
fast convergence.

To calculate ground state expectation values of an arbitrary operator $A$,
we take the large $m$ limit of
\begin{equation}
\frac{ \langle \Phi_0 | A |  \Phi_0 \rangle}{ \langle \Phi_0 |\Phi_0 \rangle} =
\lim_{m \rightarrow \infty} 
\frac{\langle m | A | m \rangle}{\langle m | m \rangle}
 . \label{project:Op}
\end{equation}

For large values of $m$, we cannot evaluate $H^m$ directly.
The number of position-space states
generated diverges exponentially with the power $m$,
so we calculate $H^m$ by a stochastic method similar to Neumann-Ulam
matrix inversion.\cite{negele}
We decompose $H$ into a product of a transition probability $p_{\alpha \beta}$
to make a transition from state $\alpha$ to state $\beta$ and a residual
weight $w_{\alpha \beta}$ as
\begin{equation}
H_{\alpha \beta} = p_{\alpha \beta} w_{\alpha \beta} \label{project:decomp}
\end{equation}
where
\begin{equation}
\sum_{\beta} p_{\alpha \beta} = 1, ~~~ p_{\alpha \beta} \geq 0
\label{project:condition}.
\end{equation}
To evaluate 
\begin{eqnarray} 
\langle \alpha_0 | H^m | \alpha_m \rangle =
\sum_{\alpha_1,...,\alpha_{m-1}} & &\langle \alpha_0| H | \alpha_1 \rangle 
\langle \alpha_1| H | \alpha_2 \rangle ... \nonumber \\
& &\langle \alpha_{m-1}| H | \alpha_m \rangle  
\end{eqnarray}
stochastically, we average over $m$-step random walks
$\alpha_0 \rightarrow \alpha_1 \rightarrow \cdots \rightarrow
\alpha_{m-1} \rightarrow \alpha_m$,
where each $\alpha_i$ is a position space state,
giving each walk the accumulated weight
\begin{equation}
W(\alpha_0,\alpha_1, ..., \alpha_m)
= w_{\alpha_0 \alpha_1} w_{\alpha_1 \alpha_2} \cdots
 w_{\alpha_{m-1} \alpha_m}.
\label{project:accum}
\end{equation}
The probability of the walk $\alpha_0 \rightarrow \alpha_1,\cdots 
\rightarrow \alpha_m$ is
\begin{equation}
P(\alpha_0,\alpha_1,..., \alpha_m) = 
p_{\alpha_0 \alpha_1} p_{\alpha_1 \alpha_2} \cdots
 p_{\alpha_{m-1} \alpha_m}.
\label{prob}
\end{equation}
Thus, it follows that
\begin{equation} 
\langle \alpha_0 | H^m | \alpha_m \rangle =
\sum_{\alpha_1,...,\alpha_{m-1}}
W(\alpha_0,\alpha_1, ..., \alpha_m)
\label{project:sum}
\end{equation}
for a large number of walks guided by the probability (\ref{prob}).

\subsection{Importance Sampling}
\label{project.importantsampling}
The Monte Carlo sum (\ref{project:sum}) is evaluated most efficiently
using importance sampling.
We cannot use the trial state as a guiding function for the random walk,
since the guiding function must be positive for all allowed states.
Labeling our guiding function $\Psi^G$, we let
\begin{equation} p_{\alpha \beta} = 
\frac{1}{z_\alpha} \frac{\Psi_\beta^G}{\Psi_\alpha^G} H_{\alpha \beta}
\label{project:prob}
\end{equation}
where
the normalization is simply the local energy:
\begin{equation}
z_\alpha = \sum_\beta \frac{\Psi_\beta^G}{\Psi_\alpha^G} H_{\alpha \beta} .
\label{z}
\end{equation}
Defined in this way, (\ref{project:prob}) satisfies (\ref{project:condition}),
and the residual weight is
\begin{equation}
w_{\alpha \beta} = z_\alpha \frac{\Psi_\alpha^G}{\Psi_\beta^G} ,
\end{equation}
resulting in the accumulated weight
for the $m$-step walk given by Eq. (\ref{project:accum}).

For an antisymmetric trial wave function $\Psi^T$, the standard algorithm to evaluate
\begin{equation} \langle \Psi^T | H^m | \Psi^T \rangle = 
\sum_{\alpha\beta}
	\Psi^{T*}_\alpha \langle \alpha | H^m | \beta \rangle
\Psi^T_{\beta},
\end{equation}
where $\Psi^{T*}_\alpha = \langle \Psi^T | \alpha \rangle$,
is to generate a set of $M$ initial states
$\{\alpha_i\}$ with probabilities proportional to
$Q_{\alpha_i} \propto |\Psi_{\alpha_i}^T|^2$ using Metropolis sampling
as in Variational Monte Carlo. 
At each initial state $|\alpha_i\rangle $, we start an $m$-step
random walk, ending in the state $|\beta_i\rangle $.
For large $M$,
\begin{equation}
\langle \Psi^T | H^m | \Psi^T \rangle \rightarrow \frac{1}{M}
	\sum_{i = 1}^M
\frac{\Psi_{\alpha_i}^{T\ast}  W(\alpha_i,...,\beta_i) \Psi_{\beta_i}^T}
{Q_{\alpha_i}}.
\label{project:mc}
\end{equation}

\subsection{A New Approach}
\label{project.newapproach}
The standard algorithm is inefficient since a random walk in configuration
space of length $m$ must be generated for each term in the sum
(\ref{project:mc}).
The details of the intermediate states are thrown away.
The expectation value
$\langle \Psi | H^m | \Psi \rangle $ can be evaluated more efficiently if the
generation of the initial states $\{\alpha_i\}$
is combined with the generation of the random walks.
In the random walk, new states are chosen with a probability
given by (\ref{project:prob}).
After a large number of steps, these states are distributed with a
probability 
\begin{equation}
Q_\alpha \propto z_\alpha [\Psi_\alpha^G]^2 \label{newprob}
\end{equation}
which is derived by solving the ``detailed balance'' condition
\begin{equation}
Q_\alpha p_{\alpha \beta} =
Q_\beta p_{\beta \alpha}, \label{detailed:balance}
\end{equation}
where $p_{\alpha \beta}$ is the probability to make
a transition from configuration $\alpha$ to $\beta$ given by
(\ref{project:prob}),
and $Q_\alpha$ is the probability to visit a state
$|\alpha\rangle$.\cite{binder86}
Thus
we may use states
generated in the $m$-step random walk as initial states
for new $m$-step random walks.

For maximum efficiency, we use every state generated as the starting point for
a new walk, so at each step we calculate different stages of $m$-walks
simultaneously. We simply generate one very long random walk using 
the probability (\ref{project:prob}).
At each step in the walk, we look $m$ steps into the past to evaluate an
element of (\ref{project:mc}).
The computer time needed to calculate
a given number of observations of $\langle H^m \rangle$ is independent of $m$.
An additional advantage is that since only one long random walk is generated,
we may calculate all different powers $m$ in parallel.
The fundamental observation becomes
\begin{equation}
\langle \Psi^T | H^m | \Psi^T \rangle = \frac{1}{M} \sum_{i = 1}^M
\frac{\Psi_{\alpha_{i-m}}^{T\ast} W(\alpha_{i-m},..., \alpha_i) 
\Psi_{\alpha_i}^T }
		{z_{\alpha_{i-m}} |\Psi_{\alpha_{i-m}}^G |^2 }.
\label{H_m}
\end{equation}

The method is easily generalized to evaluate
the expectation value
$A_{m} \equiv \langle \Psi | H^m A  H^m | \Psi \rangle$
for any {\em diagonal} operator $A$, such as the density or
spin structure factors,
$n_i n_j$ and $S_i^z S_j^z$.
At each stage in the walk, we look $m$ steps into the past to
obtain the expectation value of $\langle A \rangle$ and $2m$ steps
into the past to calculate the accumulated weight.
By summing $M$ observations from a walk, we find
\begin{equation}
A_m =
\frac{1}{M} \sum_{i = 1}^M
\frac { \Psi_{\alpha_{k}}^{T\ast} 
W(\alpha_{k},..., \alpha_i) \Psi_{\alpha_i}^T }
	{z_{\alpha_{k}} |\Psi_{\alpha_{k}}^G |^2 }
\langle \alpha_{j} | A |  \alpha_{j} \rangle.
\end{equation}
where $j=i-m$ and $k=i-2m$.

The speed of convergence of the procedure with power $m$
is determined by the ratio
$R = | E_1 - W | / |E_0 - W |  < 1 $, where $E_1$ is the energy
of the first excited state overlapping the trial state.
Since this gap
is caused by the finite size of the system, we generally calculated
powers of the Hamiltonian up to several times the linear system size.

\subsection{Trial Wave Functions}
\label{trial_section}
Care is needed to choose a trial state with maximal overlap
with the true ground state.
We restrict ourselves to total spin singlet states with zero momentum and
try to write a very arbitrary form yielding a good initial guess throughout the
phase diagram.

We use a Jastrow resonating-valence-bond wave function for the trial
state, written

\begin{eqnarray}
\Psi^T
& = &
\prod_{i<j,\sigma,\sigma'} f({\bf r}_{i ,\sigma} - {\bf r}_{j ,\sigma'})
|{\rm RVB} \rangle \label{trialstate} \\
& = &
\prod_{i<j,\sigma,\sigma'} f({\bf r}_{i ,\sigma} - {\bf r}_{j ,\sigma'})
P_N \prod_{\bf k}
(u_{\bf k} +
v_{\bf k} c^\dagger_{{\bf k} \uparrow} c^\dagger_{-{\bf k} \downarrow})
|0 \rangle, \nonumber
\end{eqnarray}
where $c^\dagger_{{\bf k} \sigma}$ is the usual Fermion creation operator and
$P_N$ projects the state onto the subspace with the number of particles
fixed to be $N$.

It is important that
the Jastrow factor $f$ correlate {\em all} pairs of particles independent of
spin, yielding a correlated state that is still a total spin singlet.
If we allow different Jastrow factors for like and unlike spins, we could
usually reduce the variational energy of the trial state.
However, the resultant state would be a superposition of many spin states
and in general would overlap excited states with non-zero spin {\em closer}
in energy to the ground state than the lowest spin zero excited state, resulting
in {\em slower} projection than a singlet trial state with higher variational
energy.

We write the determinantal part of the trial state in the usual
way.\cite{gros89}
The ratio $a_{\bf k} \equiv v_{\bf k}/u_{\bf k}$ is the physical quantity,
and, assuming $a_{\bf k} = a_{\bf -k}$,
we define $a({\bf r})$ as its Fourier transform:
\begin{equation}
a({\bf r}) = \sum_{\bf k} a_{\bf k} \cos({\bf k} \cdot {\bf r}).
\end{equation}
Then
\begin{eqnarray}
|{\rm RVB} \rangle & = &
P_N \prod_{\bf k}
(u_{\bf k} +
v_{\bf k} c^\dagger_{{\bf k} \uparrow} c^\dagger_{-{\bf k} \downarrow})
|0 \rangle \nonumber \\
& = &
\left( \sum_{{\bf r}_{i \uparrow} ,{\bf r}_{j \downarrow}}
a({\bf r}_{i \uparrow}-{\bf r}_{j \downarrow})
c^\dagger_{{\bf r}_{i \uparrow}} c^\dagger_{{\bf r}_{j \downarrow}}
\right) ^ {N/2} |0 \rangle
\end{eqnarray}
can be written as the
$\frac{N}{2} \times \frac{N}{2}$ determinant
\begin{equation}
| {\bf D} | =
\left| \begin{array}{cccc}
a({\bf r}_{1 \uparrow}-{\bf r}_{1 \downarrow}) &
a({\bf r}_{1 \uparrow}-{\bf r}_{2 \downarrow}) & \cdots &
a({\bf r}_{1 \uparrow}-{\bf r}_{\frac{N}{2} \downarrow}) \\
a({\bf r}_{2 \uparrow}-{\bf r}_{1 \downarrow}) &
a({\bf r}_{2 \uparrow}-{\bf r}_{2 \downarrow}) & \cdots &
a({\bf r}_{2 \uparrow}-{\bf r}_{\frac{N}{2} \downarrow}) \\
\vdots & \vdots & \ddots & \vdots \\
a({\bf r}_{\frac{N}{2} \uparrow}-{\bf r}_{1 \downarrow}) &
a({\bf r}_{\frac{N}{2} \uparrow}-{\bf r}_{2 \downarrow}) & \cdots &
a({\bf r}_{\frac{N}{2} \uparrow}-{\bf r}_{\frac{N}{2} \downarrow}) \\
\end{array} \right|
\label{det}
\end{equation}
in position space.

In this form, $|{\rm RVB} \rangle$ spans a broad class of Fermion wave functions.
A Fermi liquid state corresponds to
\begin{equation}
a_{\bf k} = \left\{ \begin{array}{ll}
1 & {\bf k} \in {\rm Fermi~sea} \\
0 & {\rm otherwise}
\end{array} \right.
\end{equation}
while by allowing other choices for $a_{\bf k}$, the wave function can describe
a pairing state, which may be s-wave, d-wave, or something more general.

\subsection{Guiding Function}
We are tempted to use the magnitude of the trial state as our guiding
wave function, but this would be a serious mistake.
By construction, the sites of a periodic lattice lie at high symmetry points,
and the nodes of a Fermion wave function also respect these symmetries.
One finds
$\langle \Psi^T_\alpha | \alpha \rangle = 0$ for a significant fraction
of states in the Hilbert space not violating the Pauli exclusion principle.
Since
\begin{equation}
\langle \Psi^T | H^p | \Psi^T \rangle = 
\sum_{\alpha_0,\ldots,\alpha_n}
\langle \Psi_{\alpha_0}^T	| H | \alpha_1				\rangle
\cdots
\langle \alpha_{m-1}			| H | \Psi_{\alpha_m}^T	\rangle
\end{equation}
where the intermediate sums over $\alpha_1,\alpha_2,\ldots,\alpha_{n-1}$
span the {\em complete} Hilbert space, we must guide with a positive function.

Every guiding function that samples the complete Hilbert space will yield
correct results. Our challenge is to pick a function that minimizes 
the statistical fluctuations of our output.
The guiding function cannot ameliorate the sign problem in (\ref{H_m}).
We can, however, choose a function to reduce the fluctuations in
$|\Psi^T|/|\Psi^G|$.
We define
\begin{equation}
\Psi^G \equiv \max \left\{ |\Psi^T|, c \Psi^B \right\},
\end{equation}
where $\Psi^B$ is a positive function, typically a good variational
state of the bosonic Hamiltonian.
We take $\Psi^B$ to be a spin-dependent Jastrow function.\cite{hellberg95}
This is similar to a choice used in continuum problems,
but on a discrete lattice, it is not necessary to match the first derivatives
as in the continuum.
\cite{caffarel88}

\begin{figure}[bt]
\epsfxsize=3.375in\centerline{\epsffile{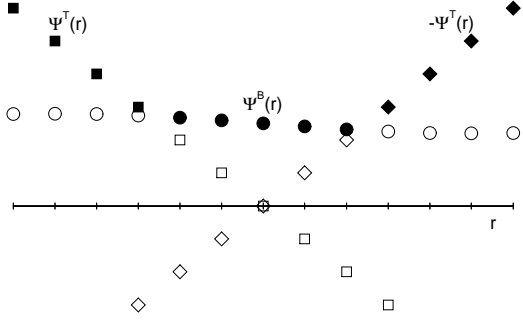}}
\caption{
Schematic behavior of the guiding function near a node.
The squares and diamonds are the trial state and its negative, respectively.
The circles are the bosonic state, and the guiding function, 
$\Psi^G = \max \left\{ |\Psi^T|, c \Psi^B \right\}$, is shown
by the filled symbols.
$r$ represents a single coordinate of one electron.
}
\label{guidewf}
\end{figure}

We rescale $c$ so the effective number of configurations contributing
to the norm is approximately $N \approx 1/L$ for an $L \times L$ system.
\cite{alexander92}
This guiding function is shown schematically in Fig.\ \ref{guidewf}.

For the guiding function,
we use the Jastrow-pairing function
\begin{equation}
\Psi^B = \prod_{i<j} f({\bf r}_{i \uparrow} - {\bf r}_{j \uparrow})
\prod_{i<j} f({\bf r}_{i \downarrow} - {\bf r}_{j \downarrow})
\prod_{i,j} g({\bf r}_{i\uparrow}  - {\bf r}_{j\downarrow} ) \label{guidestate}
\label{eq:guidewf}
\end{equation}
of Bose spin $\frac{1}{2}$ particles (i.e., two kinds of bosons, up bosons and
down bosons).
We have chosen this function to mimic the physics of the Fermion state
as closely as possible without having nodes.
Since it is not important to guide with a spin singlet function,
we use a more arbitrary spin-dependent Jastrow factor
where like spin particles are correlated differently than
opposite spin particles.

\subsection{Walking Through The Nodes}

To evaluate the determinantal wave function,
at the beginning of the random walk we calculate the determinant, 
an $O(N^3)$ operation for a $N \times N$ determinant,
and its inverse, also an $O(N^3)$ operation.
Each kinetic step in the walk changes either a row or column of the
determinant, while superexchange changes both a row and column.
Ceperley, Chester, and Kalos\cite{ceperley77} showed that the determinant
and inverse may be updated after such moves efficiently in $O(N^2)$ steps.
This so called ``inverse update'' works well for variation Monte Carlo
or fixed node Monte Carlo, but it cannot be used directly with GFMC on
a lattice since the random walk steps directly on nodes for a significant
fraction of steps.
In a reasonably dense system, we find as many as 1/3 of the steps
land on nodes.
On a node, the matrix is singular, the determinant is zero,
and the inverse is undefined.
Recalculating the determinant and inverse after walking through a node
will cause the running time to scale as $O(N^3)$.

We developed a new $O(N^2)$ technique to hop over nodes without recalculation
of the determinant or inverse.
The essence of the method is this:
When the random walk generated by the guiding function hits a state or
series of states where the determinant of the trial function vanishes,
we generate a ``detour'' walk around the
region where the matrix is singular,
rejoining the guiding walk when the determinant is non-zero again.
We stress that the real random walk goes though the node, the 
detour walk is a fictitious one which is used only to 
calculate the determinant and its inverse. It serves only 
as a calculational tool for the inverse update. 
The details of this detour-walk-approach of evaluating the
determinant and its inverse when the walk went through a node are explained 
in the Appendix \ref{inverse_up}.

Since we often land on a node
where the inverse of the matrix (\ref{det}) is not defined, 
it is difficult to calculate the
probabilities $p_{\alpha\beta}$ (Eqs.\ \ref{project:prob} and \ref{z}) for the random walk we defined earlier.
When we are on a node with $\Psi_\alpha^T=0$,
we need to choose the next step
of the walk from the various possible $\beta$ states.
Therefore we need to calculate $z_{\alpha}$ 
with Eq.\ (\ref{z}). 
Since we are on the node, calculating each $\Psi_\beta^T$ requires
a detour walk which in itself takes $N^2$ steps. Thus, the calculation 
of $z_\alpha$ is an $O(N^3)$ process, 
whereas when $\Psi_\alpha^T\ne 0$ it is an $O(N^2)$ 
process.   
Therefore, we make the following adjustments so that 
each step is an $O(N^2)$ process.
We define
\begin{eqnarray}
 p_{\alpha \beta} = \frac{1}{z_\alpha} f_{\alpha \beta}  H_{\alpha\beta}
\\
z_\alpha = \sum_\beta f_{\alpha \beta}  H_{\alpha \beta}
\end{eqnarray}

where if  $\Psi_\alpha^T \neq 0$
\begin{equation}
 f_{\alpha \beta} =
\left\{
   \begin{array}{ll}
      {\Psi_\beta^G} / {\Psi_\alpha^G}
         & \mbox{if $\Psi_\beta^T \neq 0$} \\
      {c^2 \Psi_\alpha^B \Psi_\beta^B} / {|\Psi_\alpha^G|^2}
         & \mbox{if $\Psi_\beta^T = 0$}
   \end{array}
\right.
\label{normalization}
\end{equation}

and if  $\Psi_\alpha^T = 0$,
\begin{equation}
 f_{\alpha \beta} = \frac{\Psi_\beta^B}{\Psi_\alpha^B}
\end{equation}

It is easy to show that detailed balance is obeyed by these definitions.
The advantage of using these probabilities is that if
$\Psi_\alpha^T = 0$, then calculating $\Psi_\beta^T$ for all $\beta$
with $H_{\alpha\beta} \neq 0$ not required.
Since we don't have the inverse matrix of the trial function's
determinant in state $\alpha$,
such a calculation would be computationally very expensive.

Calculating $ \Psi_\beta^B$ is always easy due to the simple form of
(\ref{eq:guidewf}), but  $ \Psi_\beta^T$ is more difficult.
The kinetic operator moves a single electron, so $ \Psi_\beta^T$
may be calculated in $O(N)$ steps since the inverse need not be updated.
The superexchange operator moves two electrons and changes both a row
and column of (\ref{det}).
Updating the inverse to calculate this term requires $O(N^2)$ steps,
which would cause the overall algorithm to scale as  $O(N^3)$ for the system.
In Appendix \ref{superexchange} we derive an $O(N)$ method of calculating
superexchange.

\subsection{Trial State Optimization}
\label{optimization}
It is important that we start the GFMC with good trial and guiding states.
In this section, we describe our method for optimizing these functions.

In continuum systems, one usually assumes a functional form for the trial
and guiding functions and optimizes a function of the energy to find the
best variational parameters.
On a lattice, there are only a finite number of distances ${\bf r}$
or equivalently wave vectors ${\bf k}$ in any given simulation, so we allow
the functions in (\ref{trialstate}) and (\ref{guidestate}) to have a parameter
describing each distance or wave vector not related by symmetry.

For the Jastrow and position space pair factors, $f({\bf r})$ and $g({\bf r})$,
we apply all rotational and mirror symmetries.
Translational symmetry is always assumed.
However, we insist only on the mirror symmetries about the axes for the
Fermion pairing field $a_{\bf k}$, so the function may be any linear combination
of an $s$ and $d_{x^2-y^2}$ pairing state.
The mirror symmetry excludes $d_{xy}$ symmetry.

For a $20 \times 20$ lattice, we have 172 parameters for the trial state
and 192 parameters for the guiding function with the pairing term.

We tried optimizing several functions of the energy, but found minimizing the
variance of the local energy to be the most robust.\cite{umrigar88}
We generate a set of configurations
$\{ \alpha _1, \alpha _2, \ldots , \alpha _m\}$ distributed according to
a weight $w _{\alpha _i}$.
The configurations remain the same throughout the minimization procedure.
We minimize the function
\begin{equation}
\sigma ^2 =
\frac{ \sum _{i=1}^m \left[ H \Psi^T_{\alpha _i}/ \Psi^T_{\alpha _i}
- E \right]^2 \left| \Psi^T_{\alpha _i} \right| ^2 / w _{\alpha _i} }
{ \sum _{i=1}^m \left| \Psi^T_{\alpha _i} \right| ^2 / w _{\alpha _i} } ,
\label{variance}
\end{equation}
where $E$ is a guess for the ground state energy that we determine self
consistently.
We use the same function to optimize both our trial and guiding functions.

With a finite random walk, the calculation of the energy in (\ref{variance})
uses many more states than the calculation of the norm.
Occasionally, this created instabilities, which we cured by
deriving another way of calculating the norm using all the neighbors in
the random walk.
We may write
\begin{eqnarray}
\langle \Psi | \Psi \rangle &=&
\sum_\alpha |\Psi_\alpha |^2 \label{oldnorm} \\
&=& \sum_\alpha
\left( |\Psi_\alpha |^2 \left(1 - A_\alpha\right) +
B_\alpha\sum_{\beta \in \{H_\alpha\}} |\Psi_\beta|^2 \right) \label{newnorm}
\end{eqnarray}
where $\{H_\alpha\}$ is the set of all states neighboring $|\alpha\rangle$ 
by application of the Hamiltonian.
We see (\ref{newnorm}) follows from (\ref{oldnorm}) if we choose $B_\alpha=C$
and $A_\alpha = C N_\alpha$ for some constant $C$, where $N_\alpha$ is the 
number of neighbors of $|\alpha\rangle$ where $\Psi$ does not vanish.
Since this version of the norm is calculated from all the states entering
the energy, no factors in the numerator of (\ref{variance}) 
are absent from the denominator.

We calculate the effective number of configurations contributing to the
normalization as
\begin{equation}
N_{\rm eff} = \left. \left( \sum_{i=1}^m
\frac{\left| \Psi^T_{\alpha_i} \right|^2 }{ w _{\alpha_i} } \right)^2 \right/
\sum_{i=1}^m \frac{\left|\Psi^T_{\alpha_i} \right|^4 }{ w _{\alpha_i}^2 } .
\end{equation}
This quantity approaches $n$ if all states contribute equally to the integral
and drops to 1 as one state begins to dominate.\cite{alexander92}
We adjust the length of our random walks so $N_{\rm eff}$ is at least 10 times
the number of parameters being optimized.

We found that close to phase separation, the standard Metropolis algorithm
develops a small acceptance ratio, and tends to stay in the same configuration
for many steps.
In order to sample more phase space quickly, we choose our configurations using
the transition probability (\ref{project:prob}) where we take $H$ to be the 
off-diagonal part of the Hamiltonian, ensuring a new configuration with each
move.
Thus the configurations are distributed according to the weight
$w_{\alpha_i} = z_{\alpha_i} | \Psi_{\alpha_i}^G|^2$, where $z_{\alpha_i}$
is given by (\ref{z}).

\section{Fitting Projection Output: Inverse Theory}

\subsection{The Ground State Energy}

The Green's Function Monte Carlo procedure takes a trial state and projects it
onto the exact ground state.
Its output consists of the observables for the energy
\begin{equation}
E^{(n)} = {\langle \Psi | H ^n | \Psi\rangle} 
\label{fit_egy}
\end{equation}
where the trial state $|\Psi \rangle$ has been normalized.
For any operator $A$ which does not commute with the Hamiltonian, using the present 
Monte Carlo method we calculate
\begin{equation}
a_{mn} = {\langle \Psi | H ^m A H ^n | \Psi\rangle}
\label{fit_corr}
\end{equation}
as functions of powers of the Hamiltonian $m$ and $n$.

These values converge to their ground state values,
except for a normalization factor, for large powers $n$ and $m$.
However, their statistical errors increase exponentially
with increasing power due to the Fermion sign problem.

To extract the most information on the ground state, we use the calculated
observable for all powers less than some maximum power $p_{max}$.
By including the highly converged small powers in the approach, 
we obtain much more
accurate ground state properties than can be obtained using the 
large powers alone. Let us consider the ground state energy as an
example to demonstrate the approach next. 

Let us define the T-spectral function $c({\cal E})$ with respect to the trial 
state $|\Psi \rangle$ as
\begin{eqnarray}
c({\cal E}) &\equiv& { 1 \over {\pi}} \lim_{\eta \to 0^+} 
Im \langle \Psi | { 1 \over {\hat H - {\cal E} + i \eta}} 
| \Psi \rangle \nonumber \\
&=& \sum_i |\langle \Psi | \Phi_i \rangle|^2 \delta ({\cal E}-E_i)
\label{spectral1}
\end{eqnarray}
To show this, one may expand the trial state in the exact eigenstates 
$|\Phi_i \rangle$  of $\hat H$ as
\begin{equation}
|\Psi\rangle = \sum_{i=0} a_i | \Phi_i \rangle.
\end{equation}
It is immediately evident from the above that the poles of $c({\cal E})$ and
of the exact spectral function are at the same energy values for those 
eigenstates $|\Phi_i\rangle$ which have non-zero overlap with the trial 
state $|\Psi \rangle$. 

In order to proceed we discretize the energy interval using a fine mesh 
with $\Delta { E}$, thus, the energy takes discrete  values ${
E^*}_m$, $m=1,2,...,M$ and the T-spectral function $c^*({ E})$ is
written as
\begin{eqnarray}
c^*({E})= \sum_{m=1}^M c^*_m \delta ({ E}-{E^*}_m)
\label{spectral2}
\end{eqnarray}
where $c^*_m \ge 0$ are  non-negative real numbers. Thus, this spectral function
is thought of as a histogram, where in each fine slice of the histogram the 
value of the integral of $c({E})$ multiplied by any function 
$f({E})$ is simply $c^*_m  f({E^*}_m)$. We have used a mesh interval
$\Delta { E}$ smaller than the finite-size gap between the lowest
and the first excited state. Thus, the contribution of the ground state 
to $c^*$ is accurately
represented as a single delta-function peak. Namely, up to some $m=m_0$,
$c^*_{m<m_0} = 0$,  while $c^*_{m_0}>0$ and $c^*_{m_0<m<m_1}=0$ and 
$c^*_{m_1}> 0$, etc. 

Then the moments of the T-spectral function $c^*(E)$ can be calculated 
using both Eq. (\ref{spectral1}) and Eq. (\ref{spectral2}) and, thus, we
obtain:
\begin{equation}
{\langle \Psi | H ^n | \Psi\rangle} 
= \sum_{m=1}^M c^*_m {E^*}_m^n  
\label{moments}
\end{equation}
where $n=0,1,...,p_{max}$. Since $p_{max} < M$, because typically
$p_{max} = 40-60$ and $M=200-500$, we have more unknowns (the $c^*_m$'s)
than equations. However, the solution needs to satisfy the constraint 
$c^*_m \ge 0$ which limits the possible solutions. The optimal way to find
the most likely solution is to minimize the $\chi^2$.

\begin{figure}[bt]
\epsfxsize=3.375in\centerline{\epsffile{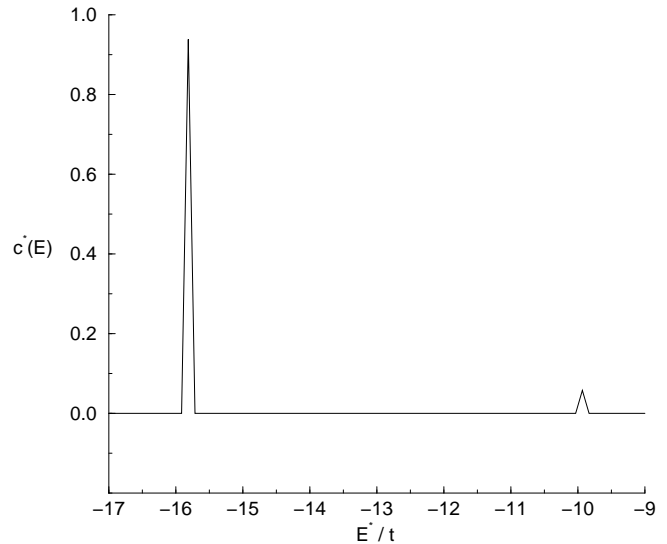}}
\caption{The T-spectral function of the full Hamiltonian.
The lowest energy peak is at the lowest energy eigenstate of the
system which is non-orthogonal to the trial state.
The value of spectral weight
is the square of the overlap of the true ground state to the initial
trial state.
}
\label{spec-energy}
\end{figure}

We gain very large computational savings by calculating all powers of 
$H$ in parallel. However, this results in statistical correlations 
between results of different
powers which must be treated accordingly.\cite{toussaint89}

We divide the measurements into $M$ bins.
The covariance matrix is defined
\begin{equation}
C_{ij} = \frac{1}{M-1}
\left( \frac{1}{M} \sum_{k=1}^M (E^{(i)}_{k}-\bar{E}^{(i)})
(E^{(j)}_{k}-\bar{E}^{(j)}) \right)
\end{equation}
where $E^{(i)}_{k}$ is the average of the $i$'th power in the $k$'th bin.
For uncorrelated output, ${\rm C}$ is diagonal.
With correlations, $\chi^2$ is defined
\begin{equation}
\chi^2 = \sum_{ij}(\bar{E}^{(i)} - E^{(i)*}) {\rm C}^{-1}_{ij} 
(\bar{E}^{(j)} - E^{(j)*}) ,
\label{chi}
\end{equation}
where $E^{(i)*}$ is the fitting function given in terms of the
coefficients $c_m$ (which are to be determined by this minimization) by
means of Eq. (\ref{moments}).

When ${\rm C}$ is diagonal, its inverse is trivial.
For more general ${\rm C}$, small errors in its components can result in large
errors in its inverse, so it is important to calculate ${\rm C}$ accurately.
Increasing the number of bins decreases the statistical error in ${\rm C}$
but increases the systematic error due to autocorrelations.
To balance these two sources of error, we choose the number of measurements
in each bin to be $n = M p_{\rm max}$, where $p_{\rm max}$ is the maximum
power of the Hamiltonian.\cite{toussaint89}
We calculate statistical errors with the bootstrap method.\cite{efron86}

\begin{figure}[bt]
\epsfxsize=3.375in\centerline{\epsffile{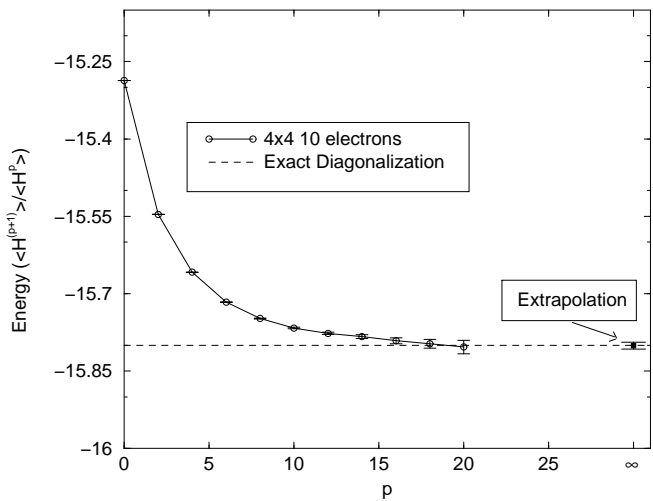}}
\caption{ The energy for 10 electrons in a $4\times 4$ lattice as
a function of the power of the Hamiltonian.
The value of the energy obtained by the extrapolation method described
in this section is also shown.  
}
\label{convergence-energy}
\end{figure}

Fig. \ref{spec-energy} gives a typical example for the T-spectral function $c^*(E)$ 
obtained from the calculation  of the two-dimensional {\em t-J} model.
The lowest value of $E^*$ where we have a delta function peak gives the
lowest eigenstate of $H$ which is not orthogonal to the trial state.
The value of the peak gives the square of the overlap of the lowest energy state to 
the trial state.

We have tested our method by comparing our results with exact results
for the $4 \times 4$ size lattice with several electrons. In Fig. \ref{convergence-energy}
we show the results for the energy as a function of the iteration
for the case of 10 electrons.

\begin{figure}[bt]
\epsfxsize=3.375in\centerline{\epsffile{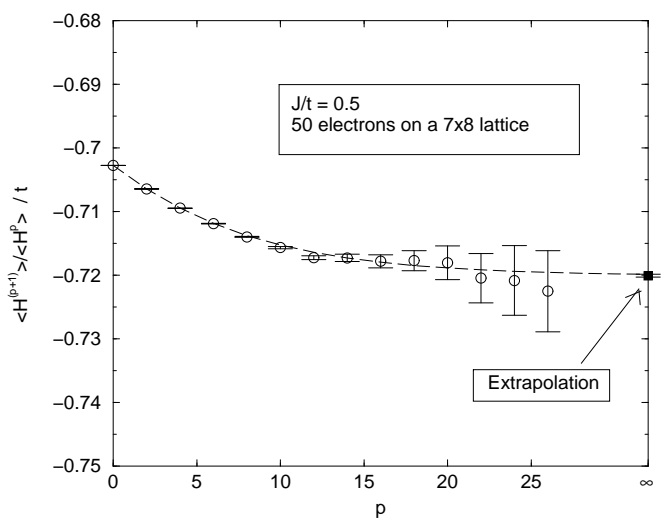}}
\caption{ The energy for 50 electrons in a $7\times 8$ lattice as
a function of the power of the Hamiltonian.
The value of the energy obtained by the extrapolation method described
in previous section is also shown.  
}
\label{c50on7x8}
\end{figure}

The energy estimate defined by
\begin{eqnarray}
E_p = {{\langle \Psi | H^{p+1} | \Psi \rangle } \over {\langle \Psi | H^p | \Psi \rangle}}
\end{eqnarray}
as a function of $p$ is shown in Fig.\ (\ref{convergence-energy}) 
starting from the Jastrow-RVB
value of the energy at $m=0$. Notice that with the
length of our walk in configuration space which we used for this
calculation, the error (which always grows exponentially) becomes 
annoyingly big for values of $p$  not shown.
The value of the energy obtained by the extrapolation method described
in previous section is also shown. By using the
information contained in all the powers of $H$ up to $p_{max}$ instead of
the just the estimates of the energy at or just below $p_{max}$ we obtain
a much better estimate for the energy.

\subsection{Other Operators}

For an arbitrary operator A, we have 
\begin{eqnarray}
{\langle \Psi | H ^n A H^n | \Psi\rangle} 
& = & \sum_{ij} (E_i E_j)^P \langle \Psi | \Phi_i \rangle 
\langle \Phi_i | A | \Phi_j \rangle \langle \Phi_j | \Psi \rangle \nonumber \\
& = & \int dE E^{2p} a(E)
\label{operator-a}
\end{eqnarray}
where the operator overlap function $a(E)$ is given by
\begin{eqnarray}
a(E) =  \sum_{ij} \delta(E+ \sqrt{E_i E_j} ) \langle \Psi | \Phi_i \rangle 
\langle \Phi_i | A | \Phi_j \rangle \langle \Phi_j | \Psi \rangle.
\label{spectral-a}
\end{eqnarray}
Following the approach for the energy T-spectral function, we can define a 
discrete overlap function
\begin{eqnarray}
a^*(E) =  \sum_{i=0}^M \delta(E- {E^*_i} ) a^*_i.
\label{discrete-spectral-a}
\end{eqnarray}
Here the values of $E$ where the T-spectral function $a^*(E)$ attains peaks are
all possible geometric means $\sqrt{E_iE_j}$ of all the eigenenergies which correspond to 
eigenstates which have non-zero overlap with $| \Psi \rangle$ and they give non-zero
matrix element of $A$. The lowest energy peak corresponds to the geometric mean of 
the ground state energy with itself, i.e $E_0 = \sqrt{E_0 E_0}$ thus, if the ground state
is not degenerate it is uniquely specified.
Here we also need to solve for all the $a^*_i$ given that they obey the following $p_{\rm max}$ 
equations:
\begin{eqnarray}
{\langle \Psi |  A | \Psi\rangle} & = & \sum_{i=0}^M a^*_i \nonumber \\
{\langle \Psi | H A H | \Psi\rangle} & = & \sum_{i=0}^M (E_i^*)^2 a^*_i \nonumber \\
\vdots \nonumber \\
{\langle \Psi | H^p A H^p | \Psi\rangle} & = & \sum_{i=0}^M (E_i^*)^{2p} a^*_i.
\end{eqnarray}

\begin{figure}[bt]
\epsfxsize=3.375in\centerline{\epsffile{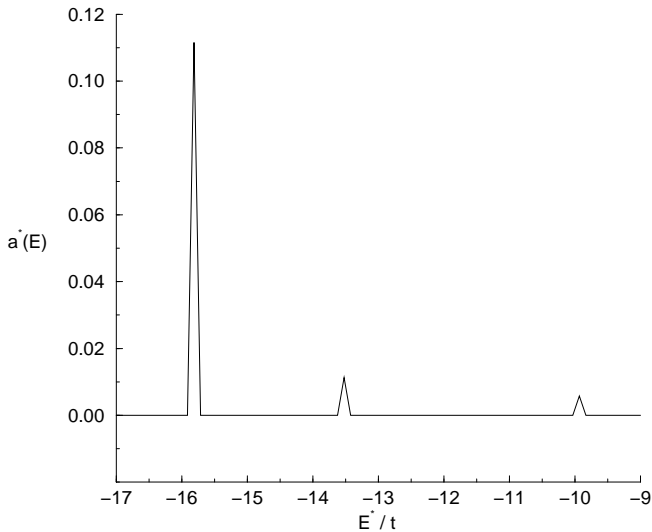}}
\caption{The T-spectral function associated with the spin-structure function.
Notice that the lowest energy peak is at the same energy as that
of the energy T-spectral function. The value of spectral weight
is related to the spin-structure function in a simple way.  
}
\label{spec-corr}
\end{figure}

\begin{figure}[htp]
\epsfxsize=3.375in\centerline{\epsffile{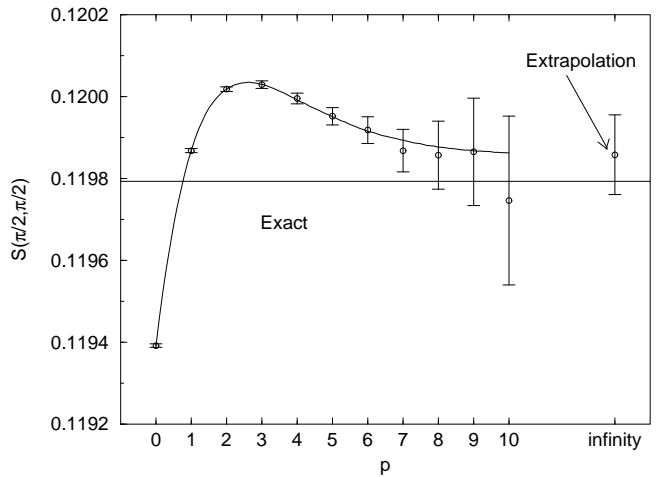}}
\caption{The spin-spin correlation function for 10 electrons in a $4\times 4$ lattice as
a function of the power of the Hamiltonian.
The value of the correlation obtained by the extrapolation method described
in previous section is also shown.
}
\label{convergence-corr}
\end{figure}

Fig. \ref{spec-corr} gives a typical example for the spectral function $a^*(E)$ 
obtained for the spin-structure function $S(\pi/2,\pi/2)$ from the calculation  
of the two-dimensional the {\em t-J} model. Notice that the energy of the lowest
peak is the energy of the lowest energy state having non-zero overlap with 
the trial state and which has non-zero matrix elements with the
operator $A$. For the value of the peak, which is $a^*_i$, the 
expectation value of $A$ can be calculated in a straightforward manner.  

In principle, we can also extract information about the excited states
along with the ground state. This possibility is indicated by the fact that
we can see higher energy peaks in these spectral functions.

In Fig.\ \ref{convergence-corr} we compare our results for the spin-spin
correlation function with that obtained by exact diagonalization
of the $4 \times 4$ size system with 10 electrons.
Notice the fine scale used to be able to distinguish
the difference between
exact and extrapolated correlation functions.

\section{Results for the 2D \mbox{\em t-J} Model}

\subsection{Maxwell Construction:  Phase Separation}

A number of methods to determine the phase separation
boundary numerically have been used.
In one dimensional systems the divergence of the density structure factor at 
long wavelengths has been used.\cite{hellberg93}
Divergence of the compressibility as determined from the second derivative
of the energy with respect to electron or hole density was
used successfully on the one-dimensional
\mbox{\em t-J} model by calculating the energy for three differing
densities.\cite{ogata91}
In the one-dimensional model, phase separation occurs between two regimes,
one with no electrons while the other contains some electrons and some holes.
For a finite system, electrons may tunnel through the vacuum, lowering the
ground state energy.
For this reason, the inverse compressibility actually passes through zero
and becomes slightly negative.
This effect is a surface effect and vanishes in the limit of infinite system
size.

In the one-dimensional \mbox{\em t-J} model,
the compressibility diverges continuously at the transition
point in contrast to the discontinuous transition
in two dimensions.\cite{putikka92,ogata91}
The behavior of the energy derivatives across the phase separation boundary
is discussed in Appendix \ref{energy_phase_separation_boundary}.
We have verified that in one dimension,
the Maxwell construction yields the same 
phase-separation boundary as that calculated using the inverse compressibility.

In the two-dimensional \mbox{\em t-J} model, the situation is more complicated.
The Fermi surface can change dramatically with electron density for a given
system size.
These strong shell effects make accurate comparisons of energies calculated
with different numbers of electrons impossible.

Many of the previous studies used a vanishing inverse
compressibility as the criterion for the onset of phase
separation.\cite{putikka92,poilblanc95}
The compressibility, however, is not the proper observable to
find the phase-separation boundary in the two-dimensional \mbox{\em t-J} model,
where the transition is first order. It is true that the compressibility
diverges in the region of phase separation, but it jumps discontinuously
at the boundary with the uniform phase. Numerically, this discontinuity 
is difficult to see in even large finite systems due to
the surface energy of the two coexisting phases. 
When one is in the region of the phase diagram where 
phase separation exists, then, the compressibility suffers strong finite-size
effects because of the rather large surface energy of the two coexisting
phases. 

\begin{figure}[htb]
\epsfxsize=3.375in\centerline{\epsffile{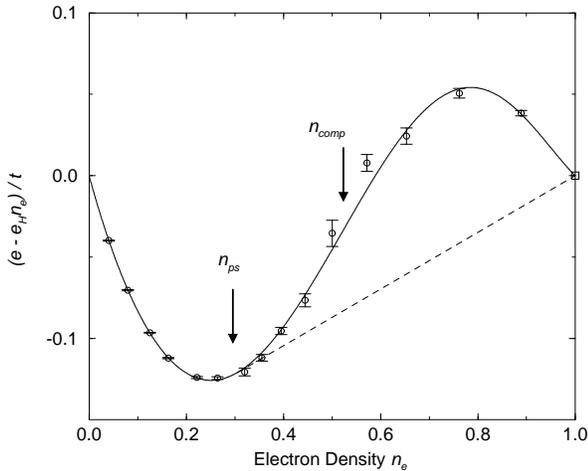}}
\caption{
The ground-state energy per site at $J = 2.5 t$ for $32$ electrons.
For clarity, the energies are shifted by a linear factor, $ - e_H n_e$.
The circles with error bars show the energies calculated on lattices
of dimensions $6 \times 6$, $7 \times 6$, ..., $28 \times 28$.
A sixth-order polynomial fit to the data is shown as the solid line,
which is extended
to
the Heisenberg energy, the square at energy zero in this shifted plot.
The dashed line shows the ground-state energy of the infinite system in the
phase-separated region.
We find the onset of phase separation occurs at
$n_{ps} = 0.296 \pm 0.004$,
while the inverse compressibility vanishes at $n_{comp} = 0.52 \pm 0.10$.
}
\label{fig:extrap}
\end{figure}

The ground-state energy as a function of electron density at $J=2.5t$ for 32
electrons on a variety of system sizes is shown in Fig.\ \ref{fig:extrap}.
These finite systems necessarily constrain the electron density to be uniform
on the length scales of the system size.
We fit the discrete data to a polynomial, $e(n_e)$, shown as the solid curve,
in order to treat the energy as a continuous function of density.
The dashed line, $e_{ps}(n_e)$, is a linear function that intersects
the Heisenberg energy, $e_H$ at electron density $n_e=1$ and intersects
$e(n_e)$ tangentially at a density labeled $n_{ps}$.

It is straight forward to show that the ground state of the infinite
system at a density $n_e>n_{ps}$ cannot be a uniform phase, because
the energy of the uniform phase, $e(n_e)$, is higher than $e_{ps}(n_e)$
at the same density.
This latter energy corresponds to the energy of a mixture of two phases,
one at electron density $n_A = 1$ and the other at electron density
$n_B = n_{ps}$.
Therefore the infinite system phase separates into two regions with densities
$n_A$ and $n_B$, and its ground-state energy is given by $e_{ps}(n_e)$,
the value of the dashed line at the average density of the system.
This is known as the Maxwell construction.\cite{emery90}

The energy of the infinite system is given by the solid line in
Fig.\ \ref{fig:extrap} for $n_e<n_{ps}$ and by the dashed line
for $n_e>n_{ps}$.
A difference between the Maxwell construction in Fig.\ \ref{fig:extrap}
and that commonly used is that the density of one of the
constituent phases, the Heisenberg phase at $n_e=1$,
lies at an extreme limit of the allowed density range.
It is not possible to add electrons to the Heisenberg solid,
which has one electron on every site,
so the dashed line is not tangent to the fitting curve
at $n_e = n_{ps}$, it is not at $n_e = 1$.
If the {\em t-J} model did allow electron densities $n_e>1$,
then the intersection point of the solid and dashed lines
would be shifted to higher densities where the curves could intersect
tangentially. The dashed line might intersect the solid curve tangentially at
the higher density point in this region.
At any electron density in the range, $n_{ps} < n_e < 1$, the system can reduce
its energy from that of the uniform phase
approximated by the fitting polynomial in Fig.\ \ref{fig:extrap},
by separating into two regions with densities $n_A = n_{ps}$ and $n_B = 1$,
resulting in an energy given by the dashed line at the average density.

In order to be stable, the energy of the infinite system must be concave
everywhere. Given the solid line in Fig.\ \ref{fig:extrap} and the 
allowed density
range of the {\em t-J} model, the dashed line drawn in the figure is
the only line possible to make the energy of the infinite system
globally concave.
This energy is given by the solid line for $n_e < n_{ps}$ and the
dashed line for $n_e > n_{ps}$.

Since the \mbox{\em t-J} model is defined only for electron densities
$0 \geq n_e \geq 1$,
the energy of the phase-separated state,
the dashed line in Fig.\ \ref{fig:extrap},
is tangent to the fitting curve at $n_e = n_{ps}$ but not at the Heisenberg
point, $n_e = 1$, an extremal density.

We never examined systems with densities $n_e \gtrsim 0.94$,
so we cannot exclude
the reentrance of a homogeneous phase in this region.
For such a phase to be stabilized,
the solid curve in Fig.\ \ref{fig:extrap} would have
to drop back below the
dashed line in this density range.
We never saw any indication of this possibility at any interaction
strength.
If a reentrant homogeneous phase did exist, the phase-separated region
would persist at densities $n_e \lesssim 0.94$.
The new Maxwell line would lie slightly below the one drawn and would
be tangent to the solid curve at both intersections, but
the phase-separated region
would persist at densities $n_e \lesssim 0.94$.
We never saw any indication of this possibility at any interaction strength.

In the infinite system, the energy is a concave function of density
with continuous first derivative.
In the phase-separated regime, the energy linearly interpolates between
the energies of the two constituent phases.
The inverse compressibility, the second derivative of the energy with respect
to density, is positive in the uniform phase and zero
where the system phase separates.
in the phase-separated regime.
On a finite lattice, the surface effects of phase separation
raise the energy, and we find the inverse compressibility
calculated from different system sizes
obtained from a fit to the discrete data
remains positive even where the system is unstable to phase separation.
For electron density $n_e$,
the completely phase-separated energy per site,
$e_{var} = e_H n_e$, is
a variational 
upper bound to the phase-separated energy.
is given by the completely phase-separated energy per site,
$e_{var} = e_H n_e$.
To show that, let  $N_e$ be the total
number of electrons on a square lattice of $N$ sites ($n_e=N_e/N<1$). 
Then, the total energy of the system is less or equal to $e_H N_e$
plus corrections which are negligible
compared to this term
in the thermodynamic limit.
To show that, let one constrain all the
electrons to be close-packed in one region of the lattice so that only the 
interaction term in (\ref{tj-ham}) is operative.
If the system chooses not to do that, it means
that the energy of any other chosen state is lower than the energy of 
this artificially imposed state. Thus, the energy per site of the 
unconstrained system is lower than $e_H n_e$.
The system is definitely unstable to phase separation
at least 
where this variational energy is 
lower than the calculated energy per site of a finite system. 

We fit the calculated energies
in the uniform phase to a low-order polynomial.
The energies in the uniform phase, where there is no surface energy,
suffer much smaller finite-size effects than energies in the
phase-separated regime.
We extrapolate the tangent of the fitting function to electron density
$n_e=1$.
At the phase-separation boundary, the tangent to the curve $e(n_e)$
extrapolates to the Heisenberg energy, $e(n_e=1) = e_H$.
This construction ensures that the resulting infinite-system energy
is concave everywhere.
This is the only possible construction that ensures the resulting
infinite-system energy is concave everywhere.
An example for $J = 2.5 t$ is shown in Fig.\ \ref{fig:extrap}.

\subsection{Results at $J = t$ }

\begin{figure}[htb]
\epsfxsize=3.375in\centerline{\epsffile{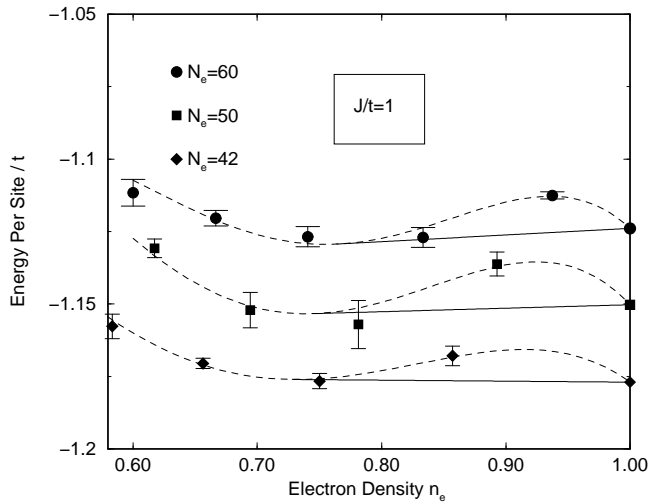}}
\caption{
The ground-state energy per site at $J = t$ for $N_e=42$ electrons and 
$N_s=49,56,64,72,81,90$ sites (bottom curve) $N_e=50$ and $N_s=56,64,72,81,90$ (second from the
bottom), 
and $N_e=60$ and $N_s=64,72,81,90,100,110,121$ sites (top curve).
Each curve has been shifted upwards with  respect to the previous by $0.025$ 
in order to distinguish them. 
 }
\label{j1}
\end{figure}

\begin{figure}[htb]
\epsfxsize=3.375in\centerline{\epsffile{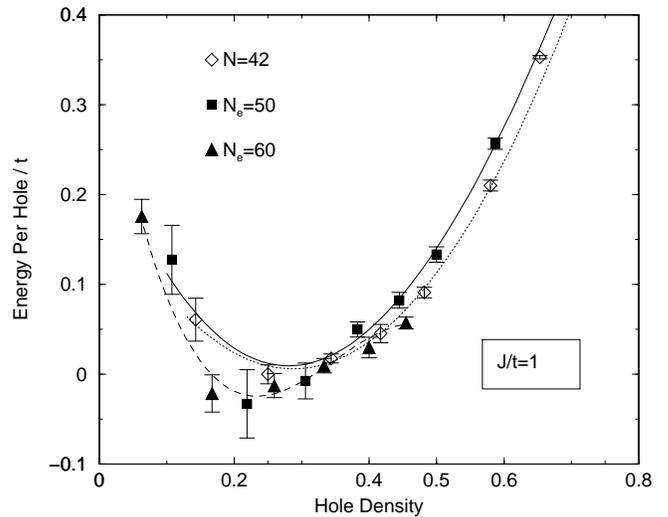}}
\caption{
The energy per hole at $J = t$ for $N_e=42$ (open diamonds) 
 $N_e=50$ (solid squares)
and $N_e=60$ (solid circles). 
The shell-effects
have non-monotonic influence on the scaling with size of the energy
per hole.
 }
\label{j1eh}
\end{figure}

Using the Maxwell construction we have determined the boundary for phase
separation of the {\em t-J} model in the $J < t$ region. In these calculations we minimize the
shell-effects when varying the electron-density by keeping the number of
electrons fixed at a closed shell configuration and changing the size of the lattice.  
In addition, we also
choose the number of electrons to correspond to a closed
shell configuration. This choice eliminates possible degeneracies of states at the
Fermi level. Such degeneracies might favor flatness of the energy as a function of
density which might be mistaken for phase separation. When one varies the number of
electrons keeping  the  system size fixed when adding an electron to a
closed shell configuration the kinetic energy goes up by a finite amount
which leads to oscillatory behavior of the energy per site versus density.
An additional (technical reason) for wanting to keep closed shell configurations is that 
on a finite lattice it leads to an energy gap between the ground
state and the first excited state which helps our projection method to converge.

In Fig.\ \ref{j1} we
study the size dependence of the critical value of electron density $n_{ps}$ for
phase separation. 
The value of $n_{ps}$ is determined 
by making a cubic polynomial fit to each curve
and using the corresponding energy for the lattice full of electrons  
calculated for the same number of electrons and the same boundary conditions.
This can be done by using the GFMC results obtained for various 
size lattices for the square-lattice spin-1/2 Heisenberg antiferromagnet 
and using the extrapolation\cite{manousakis91}
\begin{eqnarray}
E/N = e_0 + { {\lambda} N^{-3/2}} 
\end{eqnarray}
where $N$ is the total number of electrons.

\begin{table}[htb]
\caption{The energy per site for $J/t=1$ and for various electrons
densities and size lattices.
The extrapolation energy is obtained by the procedure 
described in section III-A.
The Lanczos energy is the lowest obtained by diagonalization within
a subspace of projection powers. }
\label{table:j1}
\begin{tabular}{crlll}
$N_e$ & $N_s$ & $n_e$ & $E_{Extrap}/N_s$ &  $E_{Lanc}/N_s$ \\ \hline

32 & 100 & 0.32  & -0.9019(12) & -0.9008(7) \\
   & 81 & 0.395 & -1.0204(34) & -1.0197(47) \\
   & 72 & 0.444 & -1.0767(12) & -1.0750(15) \\
   & 64 & 0.5   & -1.1249(48) & -1.1220(27) \\
   & 56 & 0.571 & -1.1727(48) & -1.1760(72) \\
   & 49 & 0.653 & -1.1828(44) & -1.1840(35) \\
   & 42 & 0.762 & -1.1853(57) & -1.1801(18) \\
   & 36 & 0.889 & -1.1637(40) & -1.1681(29) \\ \hline

42 & 121 & 0.347 & -0.9463(13) & -0.9448(5) \\
   & 100 & 0.42  & -1.0583(38) & -1.0548(39) \\
   & 81 & 0.519 & -1.1335(37) & -1.1383(41) \\
   & 72 & 0.583 & -1.1562(34) & -1.1548(15) \\
   & 64 & 0.656 & -1.1695(18) & -1.1662(22) \\
   & 56 & 0.75  & -1.1768(32) & -1.1765(24) \\
   & 49 & 0.857 & -1.1674(33) & -1.1656(16) \\ \hline

50 & 121 & 0.413 & -1.0229(36) & -1.0192(8) \\
   & 100 & 0.5   & -1.1080(38) & -1.1022(35) \\
   & 90 & 0.556 & -1.1408(49) & -1.1338(32) \\
   & 81 & 0.617 & -1.1566(39) & -1.1612(63) \\
   & 72 & 0.694 & -1.1766(58) & -1.1741(37) \\
   & 64 & 0.781 & -1.1787(89) & -1.1663(38) \\
   & 56 & 0.893 & -1.1633(38) & -1.1606(30) \\ \hline

60 & 110 & 0.545 & -1.1451(40) & -1.1442(35) \\
   & 100 & 0.6   & -1.1601(36) & -1.1602(18) \\
   & 90 & 0.667 & -1.1698(34) & -1.1697(23) \\
   & 81 & 0.741 & -1.1769(28) & -1.1759(41) \\
   & 72 & 0.833 & -1.1771(37) & -1.1710(35) \\
   & 64 & 0.938 & -1.1628(18) & -1.1612(15) 

\end{tabular}
\end{table}

The ground-state energy per site at $J = t$ for fixed number of  electrons $N_e$  and 
for various number of sites $N_s$ can be grouped together. The bottom curve in Fig.\ \ref{j1}
gives the energy per site as a function of density $n_e=N_e/N_s$ for
 $N_e=42$ and $N_s=49(7\times 7),56(7\times 8),
64(8\times 8),72(8 \times 9), 81(9\times 9)$ and $90 (9\times 10)$.
The second from the top gives the energy per site  shifted by
a constant amount of 0.025 for $N_e=50$ and $N_s=56,64,72,81,90$.
The third from the top gives the energy per site shifted by an  0.05 for
$N_e=60$ and $N_s=64,72,81,90,100,110,121$.
The unshifted energies are given in Table\ \ref{table:j1}.
If we plot these curves  on the same scale without a shift, they all fall on
almost the same curve. However, there are small deviations
from such a curve which are shell effects. 

Also shown in Table\ \ref{table:j1} is the lowest energy obtained
by diagonalizing within the subspace spanned by a subset of 2, 3,
or 4 different powers.
This is similar to the Lanczos algorithm for Quantum Monte Carlo data
introduced by Caffarel, Gadea, and Ceperley.\cite{caffarel91}
The energies obtained are upper bounds to the ground-state
energy.

We can greatly eliminate the shell effects by
examining several size lattices but keeping the number of electrons fixed.
We first fit each curve generated for fixed $N_e$ with a quartic spline.
Next from the point  $e(n_e=1)$ on the graph, 
 we construct the tangent to the spline which fits our points.  
The value of the density at which the tangent and the spline meet
gives us an estimate of the phase separation density for the given value of $J/t$.
Individually, each number of
electrons is consistent with a solution where a line beginning from the
no-hole limit ($n_e=1$) and being tangent on the polynomial (that fits the 
data points) near $n_e=0.745$ (see Fig.\ \ref{j1}).  
Therefore we conclude that the finite-size effects are small
in our method of determining $n_{ps}$ and the phase separation density
for $J/t=1$ is $n_{ps} = 0.745 \pm 0.015$.

Emery, Kivelson, and Lin\cite{emery90} calculated the phase
separation density using the energy per hole,
\begin{eqnarray}
e_h(x) = {{E(N_h) - E(0)} \over {N_h}}
\label{ehole}
\end{eqnarray}
where $E(N_h)$ is the total energy of the $N_s$-site system with $N_h$ holes,
and $x = N_h/N_s$ is the hole density.
In Fig.\ \ref{j1eh} the energy per hole is plotted
for all the points calculated for 42, 50, and 60 electrons.
The cubic fits attain minima at approximately the same values
as the tangent constructions.
The energies 
in Fig.\ \ref{j1eh} 
are not shifted
as they are in Fig.\ \ref{j1}, and the shell effects are obvious.
For each number of electrons, the energy is a smooth function of the density.
However, taking all electron numbers together, the energy is a very jagged function:
The shell effects systematically bias the energies
of systems with a given number of electrons.
Therefore it is essential to compare the energies of systems with
the same number of electrons, thus canceling
the unavoidable systematic errors.
Many previous studies of the 2D {\em t-J} model 
suffered from shell effects.
A different demonstration of shell effects is given in Ref.\ \onlinecite{hellberg98}.

\subsection{Results in the $J < t$ Region}

\begin{figure}[htb]
\epsfxsize=3.375in\centerline{\epsffile{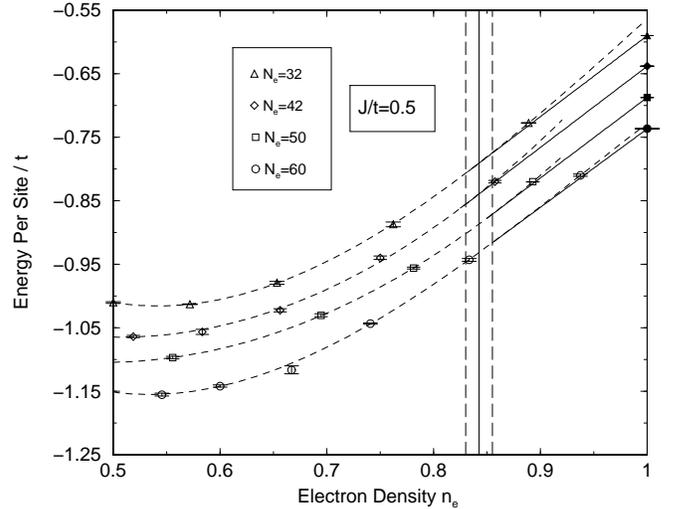}}
\caption{
The ground-state energy per site at $J = 0.5 t$ for $N_e=32$ electrons and 
$N_s=36,49,56,64,72,81,90$ sites (top curve) $N_e=42$ and $N_s=49,56,64,72,81,90$ (second from the
top), 
$N_e=50$ and $N_s=56,64,72,81,90$ (third from the top)
and $N_e=60$ and $N_s=49,56,64,72,81,90, 100, 110$ sites (bottom curve).
each curve has been shifted downward with the respect to the previous by $0.05$ 
In order to distinguish them. 
 }
\label{j05}
\end{figure}

The ground-state energy per site at $J = 0.5 t$ for fixed number of  electrons $N_e$  and 
for various number of sites $N_s$ can be grouped together. The top curve in Fig.\ \ref{j05}
gives the energy per site as a function of density $n_e=N_e/N_s$ for
 $N_e=32$ and $N_s=36(6\times 6),49(7\times 7),56(7\times 8),
64(8\times 8),72(8 \times 9), 81(9\times 9)$ and $90 (9\times 10)$.
The second from the top gives the energy per site  shifted by
a constant amount of 0.05 for $N_e=42$ and $N_s=49,56,64,72,81,90$.
The third from the top gives the energy per site shifted by an  0.1 for
$N_e=50$ and $N_s=56,64,72,81,90$.
The bottom curve gives the energy per site shifted by 0.15 for $N_e=60$
and $N_s=49,56,64,72,81,90, 100(10\times 10), 110 (11 \times 10)$. 
The unshifted energies are given in Table\ \ref{table:j05}.

\begin{table}[htb]
\caption{The energy per site for $J/t=0.5$ and for various electrons
densities and size lattices.}
\label{table:j05}
\begin{tabular}{crlll}
$N_e$ & $N_s$ & $n_e$ & $E_{extrap}/N_s$ &  $E_{lanc}/N_s$ \\ \hline

32 & 64 & 0.500 & -1.0102(19) & -1.0111(18) \\
   & 56 & 0.571 & -1.0151(24) & -1.0147(14) \\
   & 49 & 0.653 & -0.9811(27) & -0.9802(15) \\
   & 42 & 0.762 & -0.8890(33) & -0.8893(32) \\
   & 36 & 0.889 & -0.7272(11) & -0.7277(17) \\ \hline

42 & 100 & 0.420 & -0.9695(22) & -0.9698(21) \\
   & 81 & 0.519 & -1.0158(31) & -1.0143(22) \\
   & 72 & 0.583 & -1.0115(50) & -1.0030(15) \\
   & 64 & 0.656 & -0.9734(32) & -0.9734(22) \\
   & 56 & 0.750 & -0.8918(39) & -0.8851(5) \\
   & 49 & 0.857 & -0.7715(24) & -0.7733(22) \\ \hline

50 & 100 & 0.500 & -1.0015(37) & -0.9925(11) \\
   & 90 & 0.556 & -0.9975(42) & -1.0009(29) \\
   & 81 & 0.617 & -0.9898(54) & -0.9887(68) \\
   & 72 & 0.694 & -0.9310(24) & -0.9340(36) \\
   & 64 & 0.781 & -0.8594(37) & -0.8565(59) \\
   & 56 & 0.893 & -0.7232(22) & -0.7242(15) \\ \hline

52 & 100 & 0.520 & -1.0090(40) & -1.0102(44) \\
   & 90 & 0.578 & -1.0020(58) & -1.0032(80) \\
   & 81 & 0.642 & -0.9791(46) & -0.9774(42) \\
   & 72 & 0.722 & -0.9110(44) & -0.9107(62) \\
   & 64 & 0.812 & -0.8158(29) & -0.8153(20) \\
   & 56 & 0.929 & -0.6688(31) & -0.6642(6) \\ \hline

60 & 110 & 0.545 & -1.0069(33) & -1.0041(15) \\
   & 100 & 0.600 & -0.9929(24) & -0.9911(20) \\
   & 90 & 0.667 & -0.9676(32) & -0.9687(42) \\
   & 81 & 0.741 & -0.8938(11) & -0.8940(12) \\
   & 72 & 0.833 & -0.7947(31) & -0.7918(18) \\
   & 64 & 0.938 & -0.6623(25) & -0.6605(31)

\end{tabular}
\end{table}

\begin{table}[tb]
\caption{The phase separation density at $J/t=0.5$ determined
by  keeping the electron number fixed and varying the lattice size.}
\label{table:size}
\begin{tabular}{lll}
  n    &   $\sigma$ & $N_e$ \\ \hline
0.831782 & 0.00309639 & 32 \\
 0.838086 & 0.0136867  & 42 \\
 0.840684 & 0.00238198 & 50 \\
0.847915 & 0.00233589 & 52 \\
 0.857529 & 0.00913804 & 60 
\end{tabular}
\end{table}
Here again, we can greatly eliminate the shell effects by
examining several size lattices but keeping the number of electrons fixed.
We first fit each curve generated for fixed $N_e$ with a cubic spline where the
Heisenberg point has been excluded from the fit.
Next we find the point  $(n_e=1,e_H(N_e))$ on the graph, where $e_H(N_e)$ is the energy 
per electron for the Heisenberg antiferromagnet calculated on a finite-size
system with the same number of electrons $N_e$ (as discussed previously). 
Next we construct the tangent to the spline which fits our points.  
The value of the density at which the line is tangent to the spline
gives us an estimate of the phase separation density for this value of $J/t$.
These values extracted from the different sets of energies which correspond to the
same number of electrons are given in Table \ref{table:size}.
Individually, each number of
electrons is consistent with a value of $n_{ps}$ near $n_e=0.84$ (see Fig.\ \ref{j05}).  
Clearly the 42
electron data doesn't prove that there is a clear tangent at this value of $n_e$, 
but the data is consistent with this value.
Therefore we conclude that the finite-size effects are small
in our method of determining $n_{ps}$ and the phase separation density
for $J/t=0.5$ is $n_{ps} = 0.843 \pm 0.015$.

\begin{figure}[htb]
\epsfxsize=3.375in\centerline{\epsffile{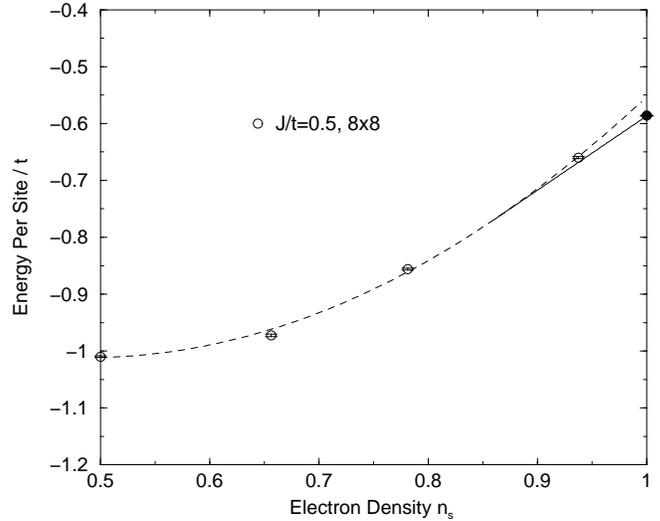}}
\caption{
The energy per site at $J = 0.5 t$ for an $8\times8$ lattice for $32,42,50$ and $60$ electrons.
}
\label{8x8}
\end{figure}

\begin{figure}[htb]
\epsfxsize=3.375in\centerline{\epsffile{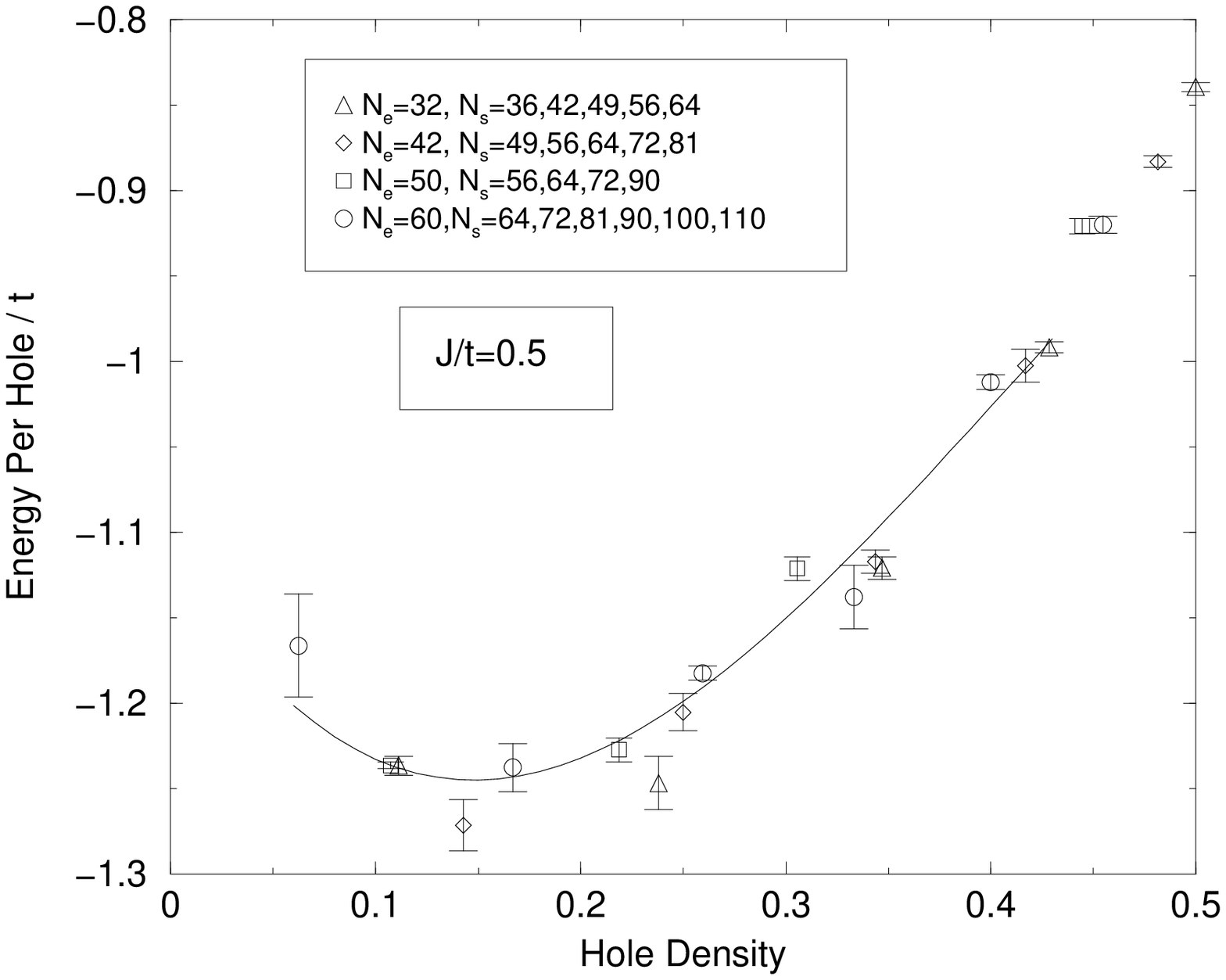}}
\caption{
The energy per hole at $J = 0.5 t$ for $32,42,50$ and $60$ electrons.
}
\label{emery}
\end{figure}

We wish to demonstrate the significance of shell effects.
Let us  select from our results of Table\ \ref{table:j05}
for $J/t=0.5$ those which correspond to the same size lattice $8\times 8$ for 
$N_e=32,42,50,60$. They are shown in Fig.\ \ref{8x8}. Notice that
even though these data also give  the same phase separation density within error
bars as that determined by our method described before, the shell effects are
large. Such deviations from a smooth curve could  lead to drawing the wrong
conclusions about phase separation boundaries.

For completeness in Fig.\ \ref{emery} the energy per hole is given
for all the points calculated for 32, 42, 50 and 60 electrons.
The curve attains a minimum at approximately the same value
as that determined by the tangent construction at the  cubic 
polynomial fit of the energy per size for a given number of
electrons. Notice, again, the shell effects.

\subsection{Results near $J^B_c$.}

Boninsegni and Manousakis\cite{boninsegni93}
(BM) found a critical value $J^B_c \simeq 0.27 t$ of 
$J/t$ below which there is no two-hole d-wave bound state.  This value of
$J^B_c$ was determined by calculating the binding energy for two holes
on lattices up to $8\times 8$. BM noticed that because the bound state wave function decays 
exponentially with distance the finite size effects were rather small. They did,
however, pursued a finite-size analysis from which they determined $J^B_c$.
Nevertheless their calculated value of the two hole binding energy at $J/t=0.7$ was
$\Delta/t=0.31(03)$ and at $J/t=0.4$ $\Delta/t=0.12(04)$. Thus, we choose the $J/t=0.3$ to examine
the question of phase separation believing that this value is very close to the 
critical value $J^B_c$.

\begin{figure}[htb]
\epsfxsize=3.375in\centerline{\epsffile{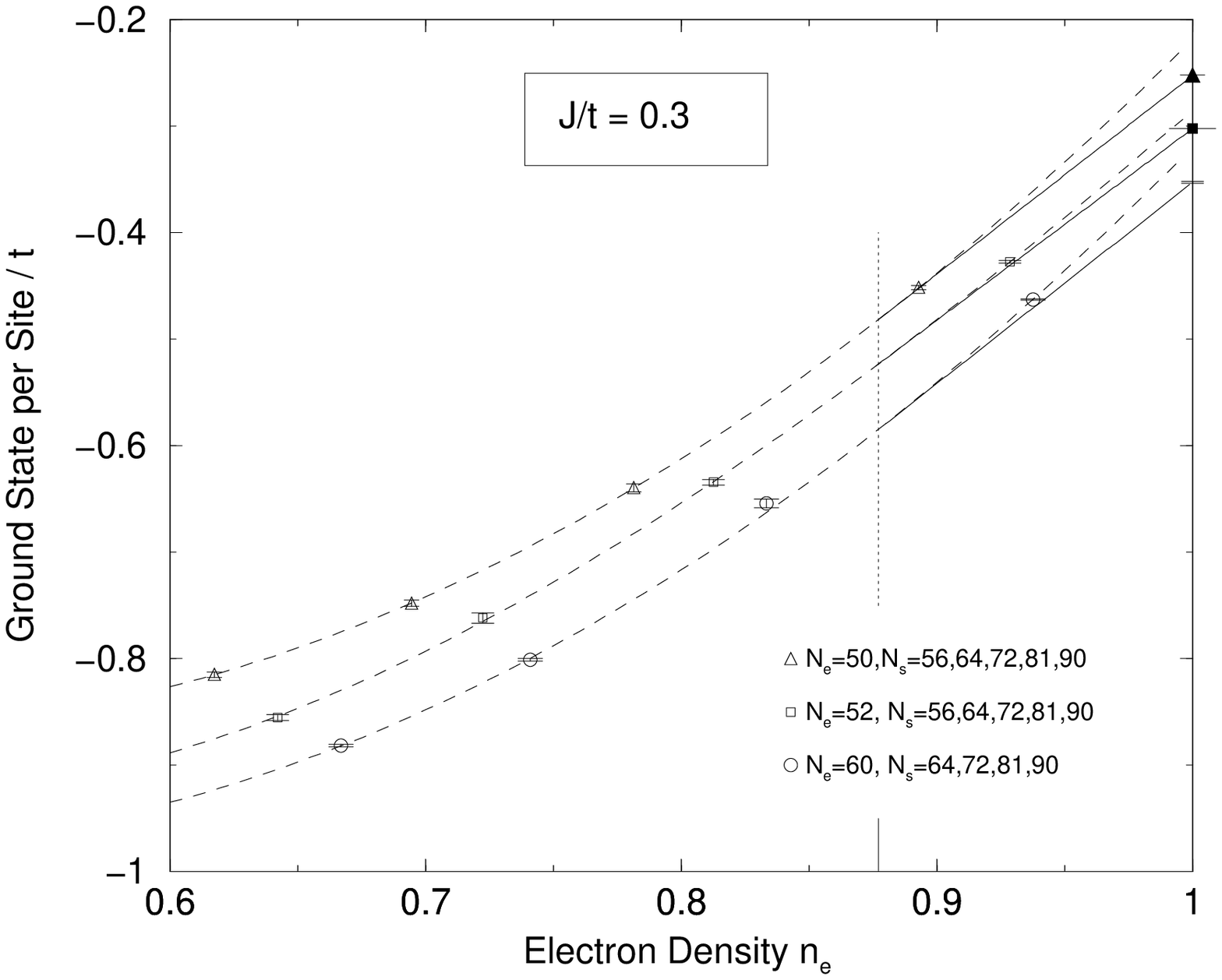}}
\caption{
The ground-state energy per site at $J = 0.3 t$ for $50,52$ and $60$ electrons
and lattices of sizes $N_s=56,64,72,81,90$, $N_s=56,64,72,81,90$ and $N_s=64,72,81,90$
respectively.
}
\label{j03}
\end{figure}

In Fig.\ \ref{j03}  we give the ground state energy as 
a function of the electron density for 50,52 and for 60 electrons
for $J/t=0.3$ as three shifted curves. 
Notice  that the values of $J_c/t$ determined from the these sets of data
are very close. We obtain: $n_e=0.877 \pm 0.010$.

\begin{table}[htb]
\caption{The energy per site for $J/t=0.3$ and for various electrons
densities and size lattices.}
\label{table:j0.3}
\begin{tabular}{crlll}
$N_e$ & $N_s$ & $n_e$ & $E_{extrap}/N_s$ &  $E_{lanc}/N_s$ \\ \hline

50 & 90 & 0.556 & -0.9494(28) & -0.9469(25) \\
   & 81 & 0.617 & -0.9199(37) & -0.9174(23) \\
   & 72 & 0.694 & -0.8469(25) & -0.8482(25) \\
   & 64 & 0.781 & -0.7381(42) & -0.7419(48) \\
   & 56 & 0.893 & -0.5510(13) & -0.5523(15) \\ \hline

52 & 90 & 0.578 & -0.9407(31) & -0.9372(38) \\
   & 81 & 0.642 & -0.9064(41) & -0.9099(48) \\
   & 72 & 0.722 & -0.8110(38) & -0.8095(29) \\
   & 64 & 0.812 & -0.6881(34) & -0.6852(49) \\
   & 56 & 0.929 & -0.4806(18) & -0.4777(9) \\ \hline

60 & 90 & 0.667 & -0.8814(11) & -0.8815(13) \\
   & 81 & 0.741 & -0.8010(13) & -0.7976(23) \\
   & 72 & 0.833 & -0.6538(34) & -0.6496(13) \\
   & 64 & 0.938 & -0.4640(23) & -0.4615(9)

\end{tabular}
\end{table}
\begin{figure}[htb]
\epsfxsize=3.375in\centerline{\epsffile{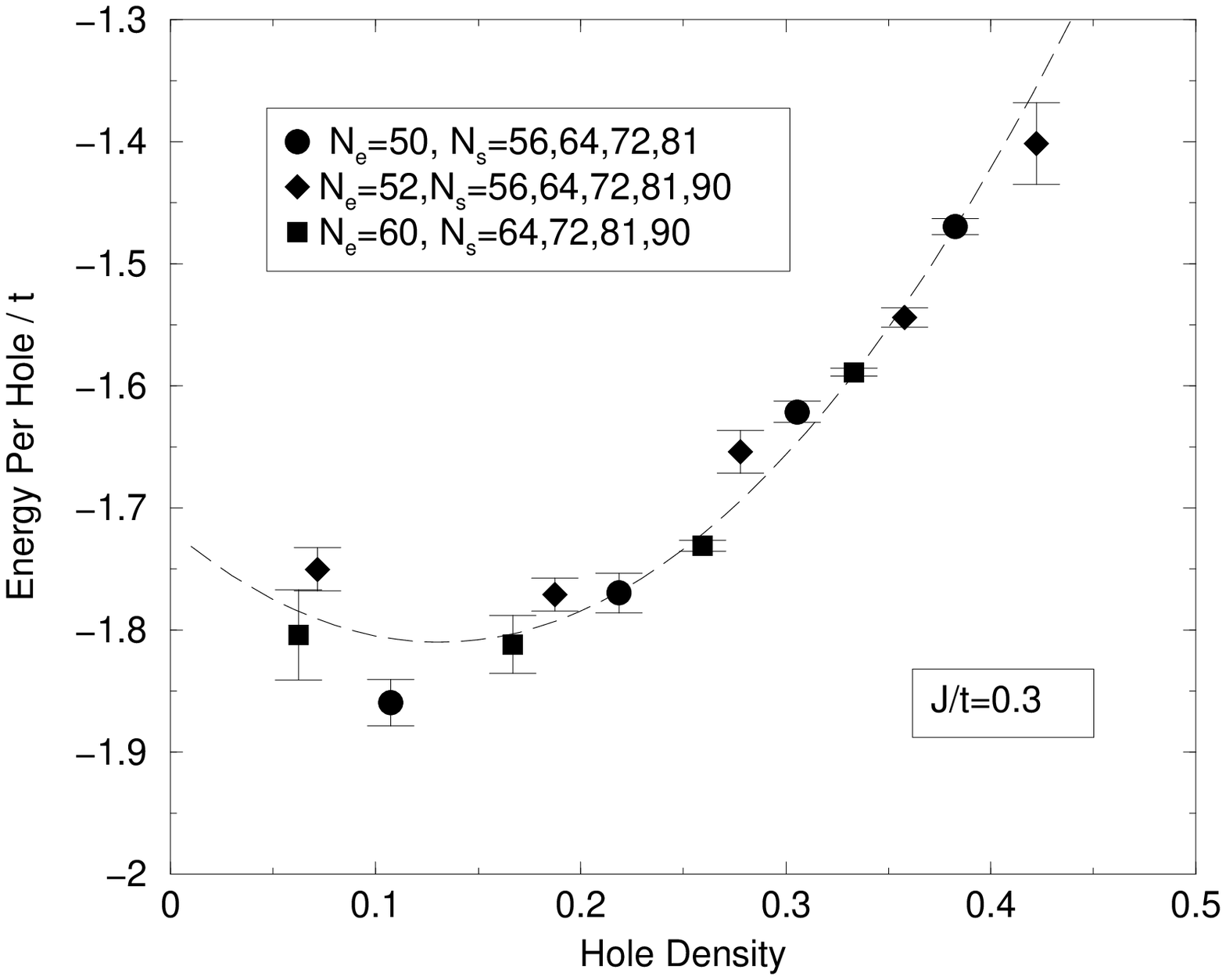}}
\caption{
The  energy per hole at $J = 0.3 t$ for $50,52$ and $60$ electrons.
}
\label{j03eh}
\end{figure}
For $J/t=0.3$ in Fig.\ \ref{j03eh} there is a minimum at $n_{ps}=0.12$ which agrees
very well with the value obtained from the tangent construction.

There are no published GFMC results for the two-hole case for 
$J/t=0.3$. We can obtain an estimate for the single-hole energy by fitting the calculated
values for that as a function of $J/t$ to a form
$E=E_0 + a J^{2/3}$.  The two-hole binding energy for $J/t=0.3$ can be estimated using the
formula which were used by Boninsegni and Manousakis\cite{boninsegni93} to 
obtain the critical value of $J^B_c$. Thus, assuming that holes are 
bound in pairs and they form a dilute gas of hole-pairs, we can obtain a value for 
the energy per hole in such a case. This value of this energy is higher than the value of the 
energy per hole at the minimum of our curve in Fig.\ \ref{j03eh}.
This is another indication that there is more binding energy gained due to the
phase separation of the pairs of holes from the electrons in an antiferromagnetically
ordered state. 
 
\subsection{Results below $J^B_c$.}

\begin{figure}[htb]
\epsfxsize=3.375in\centerline{\epsffile{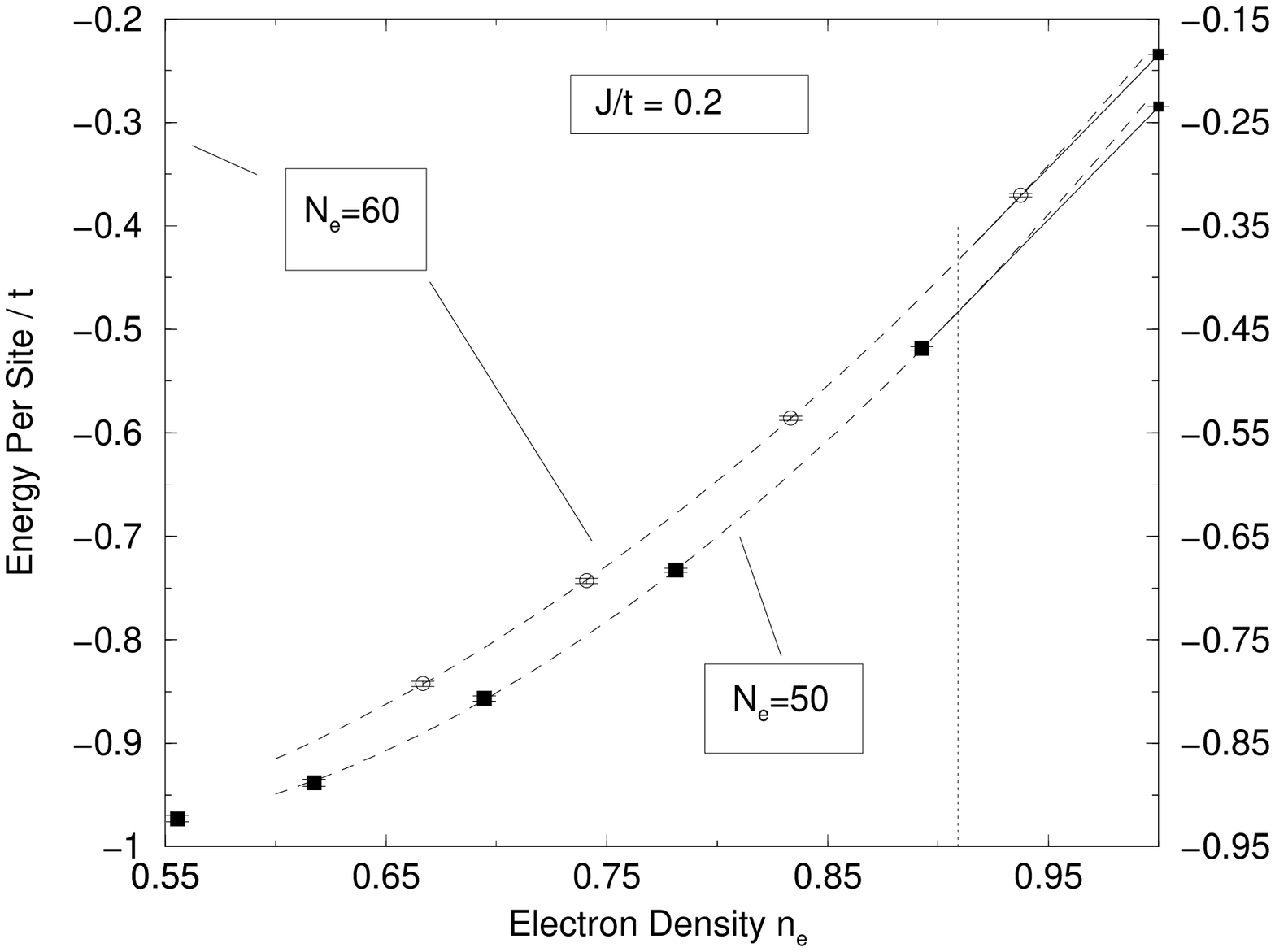}}
\caption{
The ground-state energy per site at $J = 0.2 t$ for $50$ and $60$ electrons
and lattices  with sizes $N_s=56,64,72,81,90$ and $N_s=64,72,81,90$ respectively.
}
\label{j02}
\end{figure}

Here we examine the situation below the critical value $J^B_c$ for 
two hole d-wave bound state in the 2D {\em t-J} model. We shall examine the 
energy at $J/t=0.2$. First of all the ground state energy per site for 
$N_e=50$ and $N_e=60$ and for various size lattices is shown in Fig.\ \ref{j02}
and is given in Table\ \ref{table:j02}.

Notice again that the values of $J_c/t$ determined from the two sets of data
are very close, we find: $n_{ps}=0.909 \pm 0.008$.

\begin{table}[htb]
\caption{The energy per site for $J/t=0.2$ and for various electrons
densities and size lattices.}
\label{table:j02}
\begin{tabular}{crlll}
$N_e$ & $N_s$ & $n_e$ & $E_{extrap}/N_s$ &  $E_{lanc}/N_s$ \\ \hline
50 & 90 & 0.556 & -0.9246(35) & -0.9211(54) \\
   & 81 & 0.617 & -0.8905(34) & -0.8822(15) \\
   & 72 & 0.694 & -0.8103(28) & -0.8098(32) \\
   & 64 & 0.781 & -0.6825(23) & -0.6826(27) \\
   & 56 & 0.893 & -0.4681(18) & -0.4717(23) \\ \hline

60 & 90 & 0.667 & -0.8426(32) & -0.8414(14) \\
   & 81 & 0.741 & -0.7419(24) & -0.7432(17) \\
   & 72 & 0.833 & -0.5857(23) & -0.5853(31) \\
   & 64 & 0.938 & -0.3701(17) & -0.3693(7)

\end{tabular}
\end{table}
\begin{figure}[htb]
\epsfxsize=3.375in\centerline{\epsffile{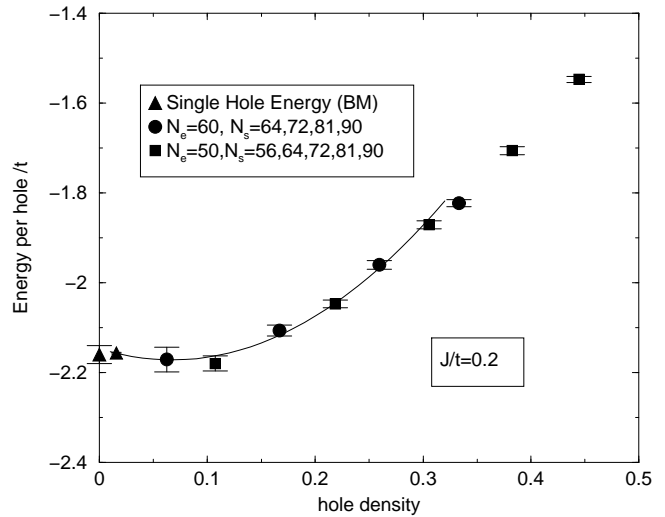}}
\caption{
The  energy per hole at $J = 0.2 t$ for $50$ and $60$ electrons.
The single-hole energy as obtained from Boninsegni and Manousakis 
is also plotted for $8\times 8$ and $10\times 10$ size lattices.
}
\label{j02eh}
\end{figure}
Below $J^B_c$ where there is no two-hole d-wave bound state, (assuming there are no bound states
in other channels) if there is no phase separation the minimum energy per hole should 
be the single-hole energy at zero hole density. At $J/t=0.2$ the single-hole
energy was  also calculated by Boninsegni and Manousakis 
for an $8\times 8$ and $10\times 10$ size lattices. This value of the energy
is shown if Fig.\ \ref{j02eh} and it is clearly higher than the minimum of the
energy per hole curve which occurs at approximately hole density of $x=0.09$.

\section{Phase Diagram of the {\em \lowercase{t}-J} Model}

\begin{figure}[htb]
\epsfxsize=3.375in\centerline{\epsffile{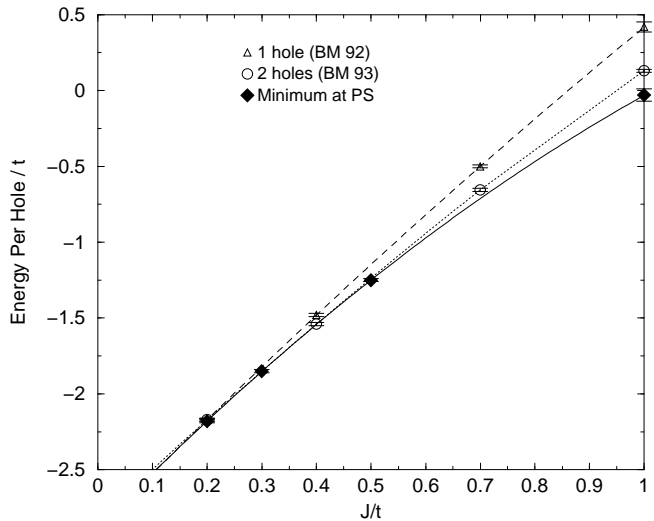}}
\caption{
The energy per hole at the density where the phase separation minimum occurs as a 
function of $J/t$ (solid line). This is compared to the energy per hole obtained
from the single hole calculation of Boninsegni and Manousakis ('92) 
(dashed line) and to the energy per hole obtained
from the 2 hole calculation of Boninsegni and Manousakis ('93).
}
\label{minimum}
\end{figure}

In Fig.\ \ref{minimum} we present as a function of $J/t$ the minimum energy per hole (solid line) 
(the minimum of the energy per hole versus density for a given value of $J/t$).
 We also plot the single hole energy
(obtained by Boninsegni and Manousakis\cite{boninsegni92}) as a function of $J/t$ which is the energy 
per hole in the case
of isolated non-interacting holes in the system (dashed line). 
In  addition, the energy per hole is compared with the
energy per hole obtained by Boninsegni and Manousakis\cite{boninsegni93} from calculation of
two holes in the {\em t-J} model. The energy per hole in this latter calculation gives the
energy per hole in the case of isolated bound hole pairs (dotted line). 
Notice that while the dashed line and the dotted line meet at $J/t \sim 0.3$,
the minimum at the phase separation density and the dotted line do not meet.
Notice that the additional energy gained to to phase separation decreases
with decreasing $J/t$ as expected.

\begin{figure}[htb]
\epsfxsize=3.375in\centerline{\epsffile{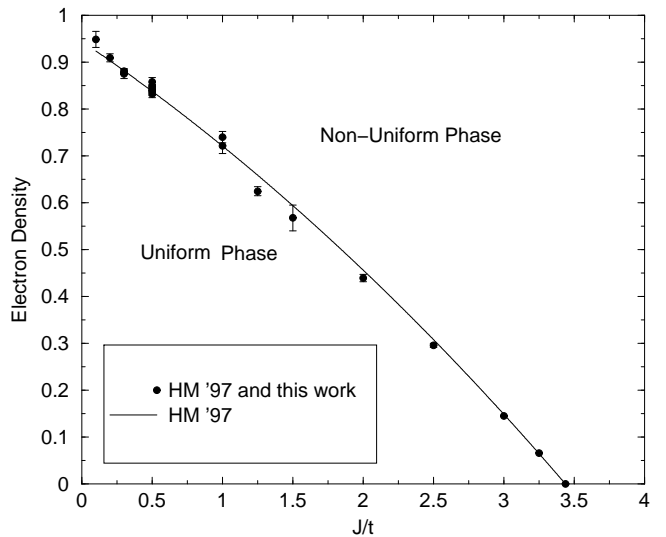}}
\caption{
The phase separation boundary as calculated using the present method
and the Maxwell construction.
}
\label{pd}
\end{figure}

In Fig.\ \ref{pd} we show the phase separation boundary obtained
for all values of $J/t$ using the method described in the present
paper and the Maxwell construction. In Table\ \ref{table:nps} we give the
phase separation boundary as determined for various values of $J/t$ from
the various size lattices and number of electrons. 

\begin{table}[htb]
\caption{The phase separation boundary as calculated using the present method.
The last value with $n=0$ is derived analytically in Ref.\ \protect\onlinecite{hellberg95}.}
\label{table:nps}
\begin{tabular}{lll}
J  &    n    &   $\sigma$  \\ \hline
0.1 & 0.9484 & 0.017   \\
0.2 & 0.909 & 0.008 \\
0.3 & 0.877 & 0.010  \\
0.5 & 0.845 & 0.015 \\
1.0 & 0.730 & 0.016 \\
1.25 & 0.624 & 0.010 \\
1.5 & 0.568 & 0.027 \\
2.0 & 0.439 & 0.008 \\
2.5 & 0.296 & 0.004 \\
3.0 & 0.145 & 0.0016 \\
3.25 & 0.0662 & 0.0006 \\
3.4367 & 0 & 0 
\end{tabular}
\end{table}

A more complete phase diagram for the 2D {\em t-J} model as a function of 
$J/t$ and doping was given in Fig. 3  of Ref.\ \onlinecite{hellberg97}. 
That phase diagram is also accurate in the low
density region where exact 
calculations can be done.\cite{hellberg95}

\section{Comparison with other calculations}

\begin{figure}[htp]
\epsfxsize=3.375in\centerline{\epsffile{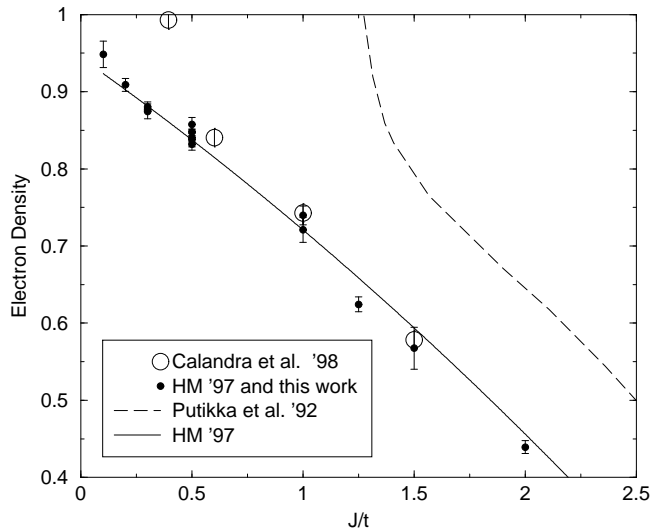}}
\caption{
Comparison of our phase-separation boundary with that of Putikka
{\em et al}.\protect\cite{putikka92} and of CBS.\protect\cite{sorella98}
}
\label{compare}
\end{figure}

In Fig.\ \ref{compare},
our phase diagram is compared to the recent fixed node Monte Carlo calculations
of Callandra, Becca, and
Sorella (CBS)\cite{sorella98}
and to the high temperature series expansion calculations of Putikka,
Luchini, and Rice.\cite{putikka92}
Notice that our phase diagram and that of CBS are very close except in 
the delicate physical region of small $J/t$. Therefore, we can draw a
relatively strong conclusion from this comparison: The findings 
drawn from the early studies of the {\em t-J} model
that the physical region of the model is safely away 
from the phase separation boundary are not
correct. What our work and the work of CBS find is that the interesting
region of $J/t$ is either next to the phase separation boundary or inside the
phase separated region. In both cases phase separation fluctuations could play 
an important role in the mechanism for superconductivity in the copper oxides.

\begin{figure}[htp]
\epsfxsize=3.375in\centerline{\epsffile{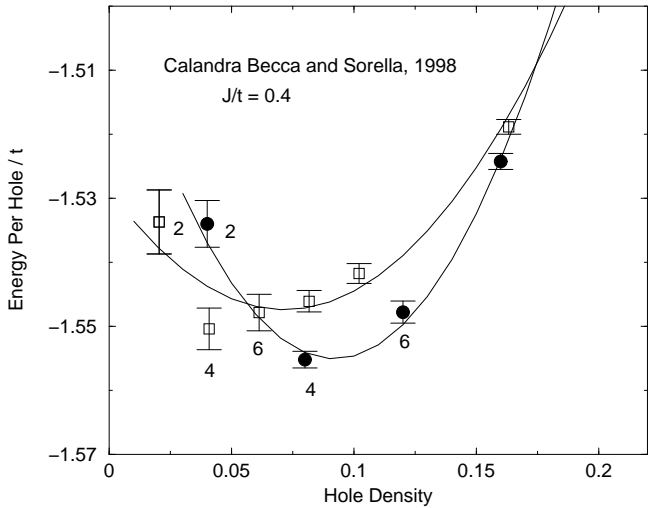}}
\caption{
Quadratic
fits of the results obtained by CBS\protect\cite{sorella98} 
for systems with 50 sites (solid circles) and
98 site (open squares) to a quadratic polynomial. The result for the lowest value of $x$
for the 98 site system was not included in the original publication by CBS.
}
\label{sorella}
\end{figure}

There is an important difference between our results and those of 
CBS. Our results indicate that phase separation
in the {\em t-J} model is present for all $J/t$, while the conclusion
of CBS is that there is a finite value of $J\simeq 0.4t$ below which 
there is no phase separation. The reason for this disagreement
is that this region requires a very high degree of accuracy in the
numerical results. We would like to discuss the results of CBS
where they claim that at $J/t=0.4$ there is no phase separation for
lattices of size $N_s=98$. In Fig.\ \ref{sorella} we plot the results of 
CBS for this value of $J/t$ for  50 sites (solid circles) and 98 sites
(open squares). The result for the lowest value of $x$
for the 98 site system was not included in the original publication by CBS.
CBS were kind enough to calculate it at our request and to communicate it to us. 
Without using that point, CBS
concluded that the fact that we found PS at $J/t=0.4$ was a finite-size
effect because the energy per site in their
largest size system had no minimum.
With the most recently calculated point for the 98-site, $e_h(x)$ has a minimum
at $x_c \approx 0.072$.
This is close to our value of
about $x_c \approx 0.1$ for $J/t=0.4$.

Let us now examine more specifically the results in Fig.\ \ref{sorella}.
We have labeled by 2, 4, and 6 the points which correspond to 
2, 4, and 6 holes in the
50 and 98 site lattices. Notice that the  energy of 4 holes is the same within
error bars in both lattices. The same is true for the 6 hole case.
Thus, the energy for 2,4 and 6 holes {\it seems} to be independent of the
size of the lattice within error bars. This can be a either a)
a genuine characteristic of presence of phase separation where 
the two, four and six-hole bubbles in a much larger system  do not
feel the size effects because they are self bound at a characteristic
size much smaller than the total system or 
b) a result of shell-effects which we have discussed and are minimized in
our calculation or c) the calculation of CBS has larger systematic or statistical errors
than  those reflected by their error bars.

White and Scalapino (WS) calculated the energy per hole on systems with
cylindrical boundary conditions, that is, systems with open boundaries in
one direction of the lattice and periodic in the 
other.\cite{white98a,white98b,white99}
They estimate the energy per hole, Eq.\ (\ref{ehole}), by comparing the energy
of a system with holes to the energy of the same system with no holes.
In Ref.\ \onlinecite{white99}, WS argue that their approach is more
accurate than that obtained by other methods simply because it gives a
lower energy per hole.
However, the energy per hole calculated in this way on systems with open
boundary conditions is not variational and, as shown below,
can artificially {\em underestimate} the energy per hole.

\begin{figure}[htb]
\epsfxsize=3.375in\centerline{\epsffile{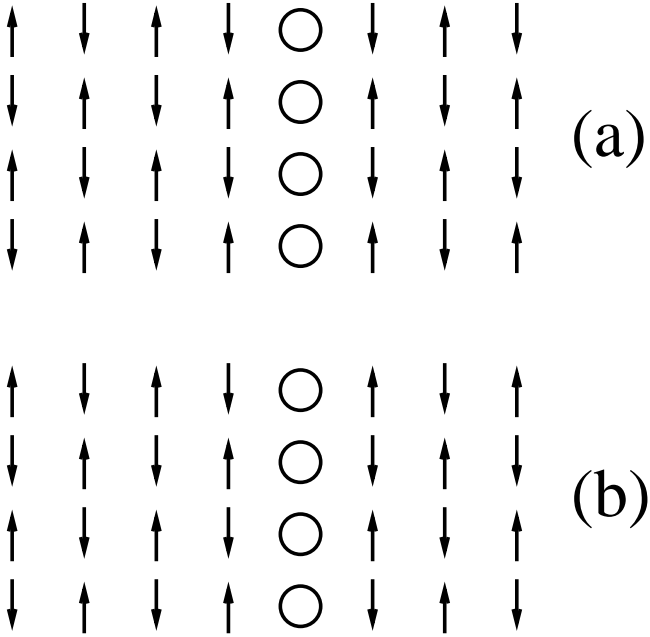}}
\vspace{.1in}
\caption{A two-dimensional stripe-type phase separated state. State (b) has a pi-phase
shift  which accommodates the hole motion along the stripe but frustrates the AF order
in the case of periodic boundary conditions. In state (b)  this twist of the order 
parameter has no magnetic energy cost with open boundary conditions along 
the x direction. Thus, periodic BC conditions in this case frustrate either
the hole motion along the ``stripe'' (state (a)) or the antiferromagnetic state
(state (b)) along the boundary bonds.
}
\label{white3}
\end{figure}

Systems with open boundary conditions can be made from fully periodic systems
by removing a row of bonds.
Clearly this process disrupts the periodic ground state and
raises the energy.\cite{hellberg99}
Both the energy of the system with $N_h$ holes, $E(N_h)$, and 
the energy of the system with no holes, $E(0)$, increase with open boundary conditions,
but generally not by the same amount.
The system with holes has more degrees of freedom
than the no-hole system, allowing it to respond more effectively
to the broken bonds.
For example, a system with holes has
freedom to twist the antiferromagnetic order
parameter at the boundary required by certain 
phase separated states. An example of such a state
is a single ``stripe'' shown in Fig.\ \ref{white3}.
In this example, a striped or structured phase separated state is stabilized
in the middle of the system.

Thus, one expects that the energy per hole obtained with open 
boundary conditions (using as a reference state
the no hole energy with open boundary conditions) can be lower than the exact 
energy per hole
obtained with periodic boundary conditions (using as a reference state
the no hole energy with periodic boundary conditions).

\begin{figure}[htb]
\epsfxsize=3.375in\centerline{\epsffile{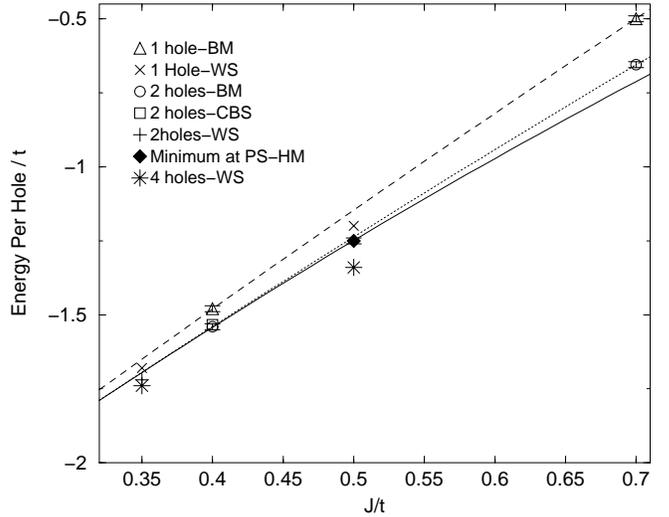}}
\caption{
Comparison of the energy per hole $e_h(N_h)$ at $N_h =$ 1, 2, and 4 holes and 
at the phase separation minimum. 
The dashed, dotted, and solid lines are polynomial fits to 
$e_h(N_h=1)$ from BM, 
$e_h(N_h=2)$ from BM, 
and $e_h$ at the phase separation minimum from HM.
Notice that because of the cylindrical boundary 
conditions which frustrates the no-hole state, 
WS tend to get more lowering of the 
energy when they introduce holes. 
For these size lattices
the finite-size effects on the 1 and 2 hole calculations are smaller
than the symbol size. For comparison we have also
placed the result of CBS for 2 holes in a 50 site lattice which is available
for $J/t=0.4$. Notice that the CBS and BM results are nearly identical.  
}
\label{white-compare}
\end{figure}

In Fig.\ \ref{white-compare}, we compare the results of various calculations
on similar size lattices for the energy per hole for 1, 2, 4 holes and at our phase separation
minimum. The results for one and two holes are taken from the work of 
Boninsegni and Manousakis (BM)\cite{boninsegni92,boninsegni93}. The finite
size effects are smaller than the size of the symbols. In addition, the 
result for 2 holes for a 50 site cluster reported by CBS\cite{sorella98} for $J/t=0.4$ is shown as an
open square.
Notice the agreement between BM and CBS (both used periodic boundary
conditions). The value for the single hole energy obtained by WS is 
systematically lower than the value for the periodic lattice.
The cylindrical boundary conditions used by WS
frustrate the no hole state, and, as a
result, the energy of the no hole state obtained by WS is much higher than that
used by HM and CBS.
WS's calculation gives a total energy of $E(0)=-35.66 $ (in units of t and here $J/t=0.5$) for the no-hole 
state on the $8\times 8$ lattice,
while for a periodic $8\times 8$ lattice the energy 
which we (and CBS) use is much lower, $E(0)=-37.56$. 
WS's total energy for 4 electrons in a
$8\times8$ lattice is $E(4)=-41.028 \pm 0.075$, while we find $E(4)=-42.23 \pm 0.12$
on a periodic lattice.
Thus WS obtain a value for the energy per hole $e_h(4)=-1.34$,
while our result corresponds to $e_h(4)=-1.17 \pm 0.03$.

Qualitatively 
similar conclusions can be drawn for the case of a single
hole. The results of BM for a single are shown by the dashed line 
in Fig.\ \ref{white-compare}.
Clearly, WS's single-hole energies are below those also. 
This lowering of the
energy can only be understood by the frustrating effect of open boundary on the 
antiferromagnetic state. 

Finally notice the very small difference between the energy per hole 
in the 2 hole case and in the 4 hole case obtained by WS at $J/t=0.35$.
They find: $e_h(2) = -1.72$ and energy per hole for a 
stripe $-1.737$ at the optimum doping of 4 holes per stripe. The difference
is very small 
and suggests that the WS striped state is 
only a manifestation of frustrated phase separation.

\section{Conclusions}

We have developed an efficient Greens Function Monte Carlo method
for fermions on a lattice that iteratively projects out the ground state
with no approximations.
Fermionic minus-sign fluctuations are controlled
by using all powers of the projection operator up to some maximum
and extrapolating to infinite power.
Starting from a good initial state allows
us to converge before the statistical errors become too large. 
This technique comes also with solutions to a number of other technical
problems such as a) enabling the guided random walk to walk through 
the nodes with an $O(N^2)$ algorithm using the idea of
a ``detour walk'' b) using a single walker 
to compute all the desired powers of the projection operator
$(H-W)^m$, where $m=0,1,...,p_{max}$, simultaneously.
 
This technique is applied to the two-dimensional {\em t-J} model
to investigate its phase diagram. It is found,
contrary to many previous studies 
that there is phase separation (PS)
at all interaction strengths of the {\em t-J} model. The signal
for phase separation is clear when one overcomes the following 
difficulties: 

First, the Maxwell construction is the cleanest and strongest
signal for PS because it suffers the least from finite-size effects.
Second the shell-effects can mask the signal because the
energy as a function of density 
of different numbers of electrons on a fixed lattice
is not a smooth curve.
The kinetic energy jumps discontinuously 
as electrons are added to successive shells.
Therefore, we have chosen to keep the
electron number fixed at a closed shell configuration and to change the 
size of the lattice.  
The number of electrons which form closed shell
configurations depends on the boundary conditions. To generate 
as many as possible ``magic numbers'' of closed shell configurations we have
used four types of boundary conditions. 
Periodic with 0 or $\pi$ phase shifts at the boundary
in each of the $\hat x$ and $\hat y$ directions. 

We find that for any value of $J/t$ the energy per site $e(n_e)$ as a function 
of electron density $n_e$ for finite size lattices does not remain a concave function 
at high electron density. 
There is a value of the density $n_{ps}(J/t)$ where a straight line 
starting from the no-hole energy per site is tangent to the curve 
$e(n_e)$ at $n_e = n_{ps}$.
While the energy $e(n_e < n_{ps})$ does 
not change significantly with system size, the energy of a finite
system in the phase separated regime $e(n_e >n_{ps})$ 
changes with system size and approaches this tangent line
in the infinite size limit. We interpret this as evidence for phase separation
at all values of $J/t$. The fact that the function $e(n_e)$ 
does not remain concave in our calculation above $n_{ps}(J/t)$ can be explained
by the energy cost of forming an interface between the two phases
in our finite system.

Our results have been compared to the most works of 
Calandra, Becca, and Sorella,\cite{sorella98}
and we find very close agreement.  These comparisons indicate that the early 
conclusions
that the critical $J_c/t$ for phase separation is far away from the
physical value of $J/t$ are largely invalid. This comparison
also indicates that $J_c/t$ is very small and may vanish.
We discuss recent comparison by White and Scalapino
(WS)\cite{white99} of our numerical results to theirs.
In that comparison, WS use the variational principle to argue 
that their results are  more accurate because the energy per hole in lower. 
However, we demonstrate that one should expect the exact energy per hole
on {\it periodic lattices}
to be higher than that obtained with the 
{\it cylindrical boundary conditions} used by WS. 
Thus on such different systems,
a lower energy cannot be used as a criterion for the 
accuracy of an approach.
In addition, we interpret the results of WS as evidence for 
phase separation and the appearance of
stripes in the {\em t-J} model as a finite-size effect.

\section{Acknowledgements}
We thank
F. Becca,
N.E. Bonesteel,
M. Calandra, 
Y.C. Chen,
R. Eder,
V.J. Emery,
S.A. Kivelson,
T.K. Lee,
P. Monthoux,
W.O. Putikka,
A.W. Sandvik,
C.T. Shih,
S. Sorella,
P.B. Stark,
and
S.R. White
for useful conversations.
This work was supported by the National Research Council
and the Office of Naval Research.
Computational work was performed at the Department of Defense
Major Shared Resource Center \mbox{ASCWP}\@.

\appendix

\section{Inverse Update through Nodes}
\label{inverse_up}

To evaluate the determinant in the trial state, we use the usual
``inverse update'' trick first applied to condensed matter
systems by Ceperley, Chester, and Kalos.\cite{ceperley77}
We calculate the determinant and inverse of the matrix (\ref{det}) together
at the start of each run, an operation taking $O(N^3)$ steps
for a $N \times N$ determinant.
Then with each single particle move in the random walk, we update the
determinant in $O(N)$ steps and the inverse in $O(N^2)$ steps.

Starting with the matrix ${\bf D}$ and its inverse ${\bf I}$, suppose we change
row $l$ of the matrix to ${D}_{lj} \rightarrow {r} _j$.
Since the inverse is the transpose of the matrix of cofactors
normalized by the determinant,
\begin{equation}
{I} _{ij} = {\rm cof}_{ji}({\bf D}) / |{\bf D}|, \label{cofactors}
\end{equation}
the
ratio of the determinant of ${\bf D}$ before and after the change is
\begin{equation}
q \equiv \frac{|{\bf D}'|}{|{\bf D}|} = \sum_j {r} _j {I} _{jl} .
\end{equation}
The new inverse matrix is given by
\begin{equation}
{I}'_{ij} = {I} _{ij} \left( 1 + \frac{1}{q} \delta_{lj} \right)
- \frac{1}{q} {I} _{il} \sum_k {r} _k {I} _{kj} , \label{inverseup}
\end{equation}
and one can easily confirm $\sum_j {D}'_{ij} {I}'_{jk} = \delta_{ik}$.
Changing one column of the matrix results in a similar update for the inverse.

The algorithm is straight forward, and has been used in many GFMC studies
in the continuum and in Variational Monte Carlo on a lattice.
{\em However, it cannot be used directly with GFMC on a lattice since the 
random walk steps directly on nodes for a significant fraction of steps.
}
When the matrix becomes singular, its inverse is undefined, and
the algorithm breaks down.

One way around this problem is to recalculate the determinant and inverse
after walking through a node.
However, in a reasonably dense system, a large faction of steps will land
on nodes, and the running time will scale as $O(N^3)$.

We developed a new $O(N^2)$ technique to hop over nodes without recalculation
of the determinant or inverse.
The essence of the method is this:
Let us suppose that the random walk visits a node; namely, the particles
were in a configuration $\vec R = (\vec r_{1\uparrow},\vec r_{2\uparrow},...,
\vec r_{N/2\uparrow},\vec r_{1\downarrow},\vec r_{2\downarrow}, ...,
\vec r_{N/2\downarrow})$  and  by moving a particle, say the first
up-spin particle from  position $\vec r_{1\uparrow}$ to $\vec r'_{1\uparrow}$
the determinant defined by Eq. 
(\ref{det}) is zero for the new configuration $\vec R' = 
(\vec r'_{1\uparrow},\vec r_{2\uparrow},...,\vec r_{N/2\uparrow},
\vec r_{1\downarrow},\vec r_{2\downarrow}, ...,\vec r_{N/2\downarrow})$. 
That is a problem for the
application of the inverse update. 
In order to move to the next
configuration, say where particle 2 is positioned at 
$\vec r'_{2\uparrow}$ and this corresponds to a new
configuration $\vec R''=(\vec r'_{1\uparrow},\vec r'_{2\uparrow},...,\vec
r_{N/2\uparrow}, \vec r_{1\downarrow},\vec r_{2\downarrow}, ..., 
\vec r_{N\downarrow})$,  we need the inverse matrix
$\bf I = I(\vec R')$ for the configuration $\vec R'$ and this does not 
exist because the determinant $D(\vec R')=0$.
The non-existence of the inverse is no problem for the physics because
all we need for computing the observables is the determinant, not the
inverse; the inverse matrix is only a tool which saves us from having to
recalculate the full determinant at each step.
However, in a reasonably dense system, a large faction of steps will land
on nodes, and the running time will scale as $O(N^3)$. We have been able
to use the inverse update technique by making a ``detour'' around the
node as follows. The real motion of the random walk was $\vec R \to \vec
R' \to \vec R''$ and because $D(\vec R')=0$ we cannot update the inverse
to find $I(\vec R')$ which we need in order to calculate $D(\vec R'')$.
However, all we need is $D(\vec R'')$ and $I(\vec R'')$ independently of
how the random walk got there. Let us consider the configuration $\vec 
R'_2 =(\vec r_{1\uparrow},\vec r'_{2\uparrow},...,\vec 
r_{N/2\uparrow}, \vec r_{1\downarrow},\vec r_{2\downarrow}, ..., 
\vec r_{N\downarrow})$, which is obtained from $\vec R$ by imagining
that we moved (without actually doing it)
particle 2 with spin-up to $\vec r'_{2 \uparrow}$ and let us assume that
$D(\vec R'_2)\ne 0$. Since $\vec R''$ can be obtained from $\vec R'_2$
by moving particle 1 to $\vec r'_{1\uparrow}$, $D(\vec R'')$ and $I(\vec
R'')$ can be obtained by imagining that the walk went through $\vec
R'_2$ to get to $\vec R''$. Thus, in order to calculate $D(\vec R'')$
and $I(\vec R'')$ we only need to calculate $D(\vec R'_2)$ and 
$I(\vec R'_2)$ which both exist.

When the random walk generated by the guiding function hits a state or
series of states where the determinant of the trial function vanishes,
we generate a ``detour'' walk around the
region where the matrix is singular,
rejoining the guiding walk when the determinant is non-zero again.

To choose the detour walk, we simply delay any move causing the determinant
to vanish and place the particle number and its future site
at the beginning of a list of moves to make.
For any subsequent move of a particle of the same spin, we try to move the
first particle in the list to that site.
If that move yields a non-zero determinant, we accept it and attempt
to move the next particle in the list in the same manner.
We repeat the process until either all moves give zero determinant or the
list is empty, in which case the true determinant is not zero.

Obviously, the procedure will not produce a non-zero determinant when the
true determinant is zero.
However, it is important to prove that the detour rejoins the guiding walk
at the first step with non-zero determinant.

We represent the rows of the matrix (\ref{det}) by
${\bf D} = \{|{\bf r}_1),|{\bf r}_2),\ldots,|{\bf r}_n)\}$
where $|\vec r_i)$ represents the row 
\begin{eqnarray}
|\vec r_i ) = (a(\vec r_{i\uparrow}-\vec r_{1\downarrow}),a(\vec
r_{i\uparrow}-\vec r_{2\downarrow}), ..., a(\vec r_{i\uparrow}-
\vec r_{N/2\downarrow}))
\end{eqnarray}
which is labeled by $\vec r_i$.
Suppose moving the first up particle to a new site, changing 
the first row label to ${\bf r}_1 \rightarrow {\bf s}$, yields a 
zero determinant.
Then
\begin{equation}
|{\bf s}) = \alpha_2 |{\bf r}_2) + \alpha_3 |{\bf r}_3) + \cdots + 
\alpha_n |{\bf r}_n ) 
\label{det_s}
\end{equation}
for some coefficients $\alpha_2,\alpha_3,\dots,\alpha_n$.
Let the next random walk step move 
the second particle, changing row ${\bf r}_2 \rightarrow {\bf t}$.
Simply by checking if the matrix
${\bf D}' = \{|{\bf t}),|{\bf r}_2),\ldots,|{\bf r}_n)\}$ has 
zero determinant, we can determine if the true matrix
${\bf D}'' = \{|{\bf s}),|{\bf t}),|{\bf r}_3),\ldots,|{\bf r}_n)\}$ 
is singular. If $|{\bf D}'| \neq 0$,
we accept the move, swap the particles, and try to move the
first particle again.
If $|{\bf D}'| = 0$,
\begin{equation}
|{\bf t}) = \beta _2 |{\bf r}_2) + \beta _3 |{\bf r}_3) + \cdots + 
\beta _n |{\bf r}_n )
\label{det_t}
\end{equation}
for certain coefficients $\beta _2,\beta _3,\dots,\beta _n$.
Combining (\ref{det_s}) and (\ref{det_t}) to eliminate ${\bf r}_2$, we
see that $|{\bf D}'| = 0$ implies $|{\bf D}''| = 0$.
Thus by simply checking single particle moves, we can verify that the
determinant of the matrix two steps away is zero.
The argument is easily generalized to any number of delayed moves.

For a Fermi liquid state, (\ref{det}) may be expanded into the product
of two Slater determinants, and this algorithm suffices as it stands.
However with a pairing trial state, this decomposition is not possible,
we must consider moves of opposite spin electrons causing the determinant to
vanish.

Suppose we find moving either the first up particle, changing the first row to
${D} _{1j} \rightarrow {r}_j$,
or the first down particle, changing the first column to
${D} _{i1} \rightarrow {c}_i$, results in a zero determinant
for matrix (\ref{det}).

If ${I} _{11} \neq 0$ we move the first row with the first element 
shifted by $1/I_{11}$, noting
\begin{equation}
|{\bf D}'| =
\left| \begin{array}{cccc}
{r}_1 + \frac{1}{ {I}_{11}} & {r}_2 & \cdots & {r}_n \\
{D} _{21} & {D} _{22} &  \cdots & {D} _{2n} \\
\vdots & \vdots & \ddots & \vdots \\
{D} _{n1} & {D} _{n2} &  \cdots & {D} _{nn} \\
\end{array} \right| = |{\bf D}| .
\label{changerow}
\end{equation}
We then try to change the first column in the standard manner.
We have artificially changed the upper left element of the determinant,
but since this element will be changed again before we finish the detour
walk, the change will not affect the true determinant.

If ${I} _{11} = 0$, this modified step is no longer possible,
and we need to prove that the true determinant,
\begin{equation}
|{\bf D}''| =
\left| \begin{array}{cccc}
{x} & {r}_2 & \cdots & {r}_n \\
{c} _{2} & {D} _{22} &  \cdots & {D} _{2n} \\
\vdots & \vdots & \ddots & \vdots \\
{c} _{n} & {D} _{n2} &  \cdots & {D} _{nn} \\
\end{array} \right| = 0 ,
\end{equation}
vanishes.
Here $x$ is the upper left element after both moves.

We know there exist coefficients $\alpha_2,\alpha_3,\dots,\alpha_n$
such that
${r} _j = \sum _{i \geq 2} \alpha _i {d} _{ij}$ for all $j$.
Since the inverse is related to the matrix of cofactors by (\ref{cofactors}),
${\rm cof} _{11} ({\bf D}) = 0$, and there are other coefficients
$\beta _2,\beta _3,\dots,\beta _n$ such that
$ 0 = \sum _{i \geq 2} \beta _i {d} _{ij}$ for all $j \geq 2$.
If $\sum _{i \geq 2} \beta _i {c} _i = 0$,
then $|{\bf D}''| = 0$ trivially.
Otherwise let $\gamma_i = \alpha_i + \lambda \beta_i$ where
\begin{equation}
\lambda = \left. \left( x - \sum _{i \geq 2} \alpha _i {c} _i \right)
\right/ \sum _{i \geq 2} \beta _i {c} _i .
\end{equation}
Then
$x = \sum _{i \geq 2} \gamma _i {c} _i $
and
${r} _j = \sum _{i \geq 2} \gamma _i d _{ij}$ for all $j \geq 2$,
so $|{\bf D}''| = 0$.

Again, this argument can be extended to any number of delayed moves.
By combining the two types of moves described in this section,
we are able to keep track of the
true determinant without recalculating the inverse from scratch.

\section{$O(n)$ Calculation of Superexchange}
\label{superexchange}

For a determinantal function, the kinetic terms in (\ref{normalization}) require $O(N)$ steps per
particle, so it scales as $O(N^2)$ for the system.
The superexchange term in the {\em t-J} model,
$\sum_{\langle ij \rangle} S^+_iS^-_j$, exchanges two particles,
changing both a row and a column of the determinant (\ref{det}).
In this section, we show how the amplitude of swapping two particles may be
calculated in $O(N)$ steps.

Suppose we swap the $m$'th up electron with the $n$'th down electron.
We will modify both row $m$ and column $n$ in the determinant.
We write the new elements as ${D} _{mj} \rightarrow {r} _j$ and
${D} _{in} \rightarrow {c} _i$.
Naturally, ${r} _n = {c} _m$.
One can show the ratio of the determinant before and after the swap is
\begin{eqnarray}
\frac{|{\bf D}'|}{|{\bf D}|} & = &
\Biggl(\sum_i {r} _i {I} _{im}\Biggl)
\Biggl(\sum_j {I} _{nj} {c} _j\Biggl) + \nonumber \\
&   & {I} _{nm}{c} _m
- {I} _{nm}\sum_{ij}{r} _i {I} _{ij} {c} _j .
\end{eqnarray}

Direct evaluation of the sum
$S = \sum_{ij}{r} _i {I} _{ij} {c} _j$ takes $O(N^2)$ per
pair of neighboring particles.
For this reason, many researchers evaluate the superexchange term only
every $N$ Monte Carlo steps.\cite{gros89}

Our trick is to evaluate $S$ once when a pair of particles become nearest
neighbors, and then to update it in $O(N)$ steps
for any move not disrupting the pair.

Suppose the $l$'th up electron moves ($l \neq m$), altering row $l$ in the
determinant (\ref{det}), so ${D} _{lj} \rightarrow {s} _j$.
The inverse $I$ is updated according to (\ref{inverseup}) and
${c} _l \rightarrow {c}'_l$  takes a new value.

We can write the new sum $S'$ in terms of the old sum and extra factors as
\begin{eqnarray}
S'
&=& \sum_{ij}{r} _i {I}'_{ij} {c}'_j \nonumber \\
&=& \sum_{ij}{r} _i
\left(
	{I} _{ij}
	( 1 + \frac{1}{\gamma_l} \delta_{lj})
	- \frac{1}{\gamma_l} {I} _{il} \sum_k {s} _k {I} _{kj}
\right)
{c}'_j \\
&=& S + \frac{1}{\gamma_l}
\Biggl( \sum_i {r} _i {I} _{il} \Biggl)
\Biggl( {c}'_l- \sum_j \gamma_j {c} _j \Biggl)
\nonumber
\end{eqnarray}
where $\gamma_j = \sum_k {s} _k {I} _{kj}$
is used in the inverse update.
This calculation requires only $O(N)$ steps, so the local superexchange
energy of the system may be evaluated in $O(N^2)$ time.

\section{Energy at the phase separation boundary}
\label{energy_phase_separation_boundary}

In the phase separated state, the \mbox{\em t-J} model separates into two phases,
one with all electrons (no holes) and the other with some electrons and some
holes.
The transition is continuous:  As J is increased in the phase separated
regime, the electron density in the low-electron-density phase decreases
while the proportion of Heisenberg phase increases.
The energy in the partially phase separated regime is simply the weighted
sum of the two constituent energies.
Specifically, the energy of the phase separated state is given by
\begin{equation}
E_{ps}(n,J) = \frac{1-n}{1-n_{ps}}		 E_u(n_{ps},J)
				+ \frac{n-n_{ps}}{1-n_{ps}} E_H J
\end{equation}
where $E_u(n,J)$ is the energy of the uniform density phase as a
function of electron density and interaction strength, $E_H J$ is the 
energy of the Heisenberg phase, and $n_{ps}(J)$ is the density of the
onset of phase separation.

Across the phase separation boundary, the energy is continuous as
is its first derivative with respect to density.
Using this fact, we can show that the derivative of the energy
in the phase separated regime with respect to $J$ is given by
\begin{equation}
\frac{\partial E_{ps}(n,J)}{\partial J} =
\frac{1-n}{1-n_{ps}}
\frac{\partial E_u(n_{ps},J)}{\partial J}
+ \frac{n-n_{ps}}{1-n_{ps}} E_H
\end{equation}
so the first derivative of the energy with respect to $J$ is
continuous at the phase separation point, $n=n_{ps}(J)$.
All terms of the form
$\frac{\partial n_{ps}(J)}{\partial J}$ are canceled from this expression.
Note that for $J > J_c$, where $J_c$ is the critical interaction strength
for complete phase separation, $n_{ps}(J \ge J_c) = 0$ and $E_u(n=0,J) = 0$.


\end{document}